\documentclass[aps,prd,onecolumn,groupedaddress,showpacs,nofootinbib,amssymb]{revtex4-2}
\usepackage[dvips]{graphicx}
\usepackage{amssymb}
\usepackage{amsmath}
\usepackage{graphicx,,color}
\usepackage{amsfonts}
\usepackage{bm}
\usepackage{cancel}
\usepackage{comment}
\usepackage{floatflt}
\usepackage{slashed}
\usepackage{appendix}
\usepackage{array}
\usepackage{tabularx}

\newcommand\be{\begin{equation}}
\newcommand\ee{\end{equation}}

\allowdisplaybreaks[4]

\begin{document}

\title{Enhancing Electroweak Baryogenesis: The Role of Dimension-Six Operators in the Real Singlet Extension of the Standard Model}
\author{Apostolos Giovanakis$^{1}$}
\affiliation{$^{1)}$Department of Physics, Aristotle University of
Thessaloniki, Thessaloniki 54124, Greece}

\tolerance=5000

\begin{abstract}

This study investigates the electroweak baryogenesis (EWBG) in the real singlet extension of the Standard Model (SM), including dimension-six operators, which couple the Higgs boson to a real singlet scalar field. We show that the electroweak phase transition is a one-step or a two-step phase transition, depending on the parameters of the singlet extension. In the majority of the parameter space, the electroweak phase transition proceeds as a two-step phase transition: an initial phase transition in the singlet sector at high temperatures, followed by a first-order phase transition in the Higgs sector. Interestingly, including the dimension-six operator, the electroweak phase transition occurs in regions of the parameter space, which were ruled out in the singlet extensions of the SM without the dimension-six operator. This is more evident for low singlet masses, which are excluded by the experimental constraints from the invisible decays of the Higgs boson. In addition, the real singlet extension explains the observed baryon asymmetry of the Universe introducing an additional source of \(CP\) violation in the SM via a second dimension-six operator, which contributes to the top-quark mass. These results provide new pathways for EWBG, expanding the viable parameter space for future collider and gravitational wave experiments, especially for low singlet masses.
\end{abstract}


\maketitle

\section{Introduction}

A wide range of fundamental problems in particle physics and cosmology have motivated physicists for decades to search for new physics. The Standard Model (SM) could be extended consistently to explain fundamental problems such as dark energy, dark matter, and the matter-antimatter asymmetry in the Universe. The matter-antimatter asymmetry in the Universe remains one of the most profound mysteries in modern cosmology and particle physics as it cannot be explained in the context of the SM. Nevertheless, numerous extensions of the SM potentially solve these drawbacks to describe a strong electroweak phase transition and an extra source of \(CP\) violation as it is successfully demonstrated by the two-Higgs doublet models \cite{Cline:1995dg, Dorsch:2016nrg, Wang:2019pet, Zhou:2020irf, Goncalves:2021egx, Biekotter:2022kgf, Morais:2019fnm}, the composite Higgs models \cite{Espinosa:2011eu, Bruggisser:2018mrt, Chala:2016ykx, Xie:2020bkl, Bian:2019kmg, Aziz:2013fga}, higher dimensional operators in the SM \cite{Delaunay:2007wb, Chala:2018ari, Bodeker:2004ws, Noble:2007kk, Huang:2016odd, Hashino:2022ghd}, and singlet scalar extensions \cite{Profumo:2007wc,Damgaard:2013kva,Ashoorioon:2009nf,OConnell:2006rsp,Gonderinger:2012rd,Profumo:2010kp,Gonderinger:2009jp,Barger:2008jx, Cheung:2012nb,Barger:2007im,Cline:2013gha,Burgess:2000yq,Kakizaki:2015wua,Chiang:2018gsn,Katz:2014bha,Espinosa:1993bs,Alanne:2014bra,Cline:2012hg,Beniwal:2017eik,Curtin:2014jma,Ghorbani:2018yfr,Ghorbani:2020xqv,Espinosa:2011ax,Espinosa:2007qk, Alves:2018jsw, Zhang:2023nrs, Jain:2017sqm, Vaskonen:2016yiu, Huang:2018aja, Jiang:2015cwa, Grzadkowski:2018nbc, Noble:2007kk}.

The simplest extension of the SM, which has been studied extensively in the past decades, is the real singlet extension \cite{Noble:2007kk, Profumo:2007wc,Damgaard:2013kva,Ashoorioon:2009nf,OConnell:2006rsp,Gonderinger:2012rd,Profumo:2010kp,Gonderinger:2009jp,Cheung:2012nb,Barger:2007im,Cline:2013gha,Burgess:2000yq,Kakizaki:2015wua,Katz:2014bha,Espinosa:1993bs,Alanne:2014bra,Cline:2012hg,Beniwal:2017eik,Curtin:2014jma,Ghorbani:2018yfr,Chiang:2018gsn,Ghorbani:2020xqv,Espinosa:2011ax,Espinosa:2007qk,Alves:2018jsw, Zhang:2023nrs, Jain:2017sqm, Vaskonen:2016yiu, Huang:2018aja}, which introduces a real singlet scalar field in the SM, denoted as \(S\). Then, the Lagrangian density of the scalar fields in the SM is written as,
\begin{equation}\label{Lag dens}
    \mathcal{L} = \left(D_{\mu} H \right)^{\dagger} D^{\mu} H + \frac{1}{2} \partial_{\mu} S \partial^{\mu} S - V(H, S),
\end{equation}
where \(H\) is the Higgs doublet in the SM. The most general renormalizable tree-level potential for the Higgs doublet and the real singlet scalar field can be cast into the form \cite{Profumo:2007wc, Gonderinger:2009jp, Zhang:2023nrs, Alanne:2014bra, Noble:2007kk},
\begin{equation}\label{gen_potential}
    V(H, S) = - \mu_H^{2} |H|^2 + \lambda_H |H|^4 - \frac{\mu^2_S}{2} S^2 + \frac{\mu_3}{3} S^3 + \frac{\lambda_S}{4} S^4 + \mu_{HS} |H|^2 S + \lambda_{HS} |H|^2 S^2.
\end{equation}
Now if the fourth and sixth terms in the tree-level potential (\ref{gen_potential}) are removed, the Lagrangian density has a \(\mathbb{Z}_2\) symmetry, under which \(S \to - S\) and all other fields remain unchanged. We are interested in investigating this spontaneous symmetry breaking in the singlet sector and its implications for the electroweak phase transition in the early Universe. Thus, if the Lagrangian density (\ref{Lag dens}) respects this \(\mathbb{Z}_2\) symmetry, the scalar potential (\ref{gen_potential}) reads 
\begin{equation}\label{Z2 potential}
    V (H, S) = - \mu_H^{2} |H|^2 + \lambda_H |H|^4 - \frac{\mu^2_S}{2} S^2 + \frac{\lambda_S}{4} S^4 + \lambda_{HS} |H|^2 S^2,
\end{equation}
Now we shall assume that the singlet scalar field is weakly coupled to the Higgs sector and the previous scalar potential could be further extended by introducing the following terms \cite{Oikonomou:2024jms},
\begin{equation}\label{high-op}
    V_{\text{HO}} (H,S) = \frac{\lambda}{M^2} |H|^2 S^4 - \frac{\kappa}{M^4} |H|^2 S^6,
\end{equation}
where \(\lambda\) and \(\kappa\) are Wilson coefficients. These higher-order operators originate from an effective field theory that is active at an energy scale \(M\). This scale will be assumed to be higher than the electroweak scale, of the order \(M= 15 - 100 \) TeV, a fact that is further motivated by the absence of new elementary particles observed in the Large Hadron Collider (LHC) since the detection of the Higgs boson. In principle, more higher-order operators with dimension \(D > 6\) could be introduced in the scalar potential, but their effect on the effective potential and the electroweak phase transition is significantly weak for valid values of the Wilson parameters. Consequently, the dimension-eight operator in the scalar potential (\ref{high-op}) is not studied further and in the following, the real singlet extension with higher-order operators will be described by
\begin{equation}\label{tree-level potential}
     V (H, S) = - \mu_H^{2} |H|^2 + \lambda_H |H|^4 - \frac{\mu^2_S}{2} S^2 + \frac{\lambda_S}{4} S^4 + \lambda_{HS} |H|^2 S^2 + \frac{\lambda}{M^2} |H|^2 S^4.
\end{equation}
The usual singlet extension (\ref{Z2 potential}), which is studied in the literature, will be investigated by taking \(\lambda = 0\) in the scalar potential (\ref{tree-level potential}). This extension has been studied in detail to explain the observed baryon asymmetry of the Universe \cite{Profumo:2007wc,Damgaard:2013kva,Ashoorioon:2009nf,OConnell:2006rsp,Gonderinger:2012rd,Profumo:2010kp,Gonderinger:2009jp, Chiang:2018gsn, Cheung:2012nb,Barger:2007im,Cline:2013gha,Burgess:2000yq,Kakizaki:2015wua,Katz:2014bha,Espinosa:1993bs,Alanne:2014bra,Cline:2012hg,Beniwal:2017eik,Curtin:2014jma,Ghorbani:2018yfr,Ghorbani:2020xqv,Espinosa:2011ax,Espinosa:2007qk,Alves:2018jsw, Zhang:2023nrs, Jain:2017sqm, Vaskonen:2016yiu, Huang:2018aja}.

Electroweak baryogenesis (EWBG) is a physical mechanism that generates an asymmetry in the density of baryons during the electroweak phase transition to explain the observed baryon asymmetry of the Universe. This asymmetry can be realized by satisfying the three Sakharov's conditions for baryogenesis \cite{Sakharov:1967dj}: i) baryon number violation, ii) \(C\) and \(CP\) violation, and iii) departure from thermal equilibrium, which requires a strong first-order phase transition as the Universe cools down. Although the SM contains all the essential ingredients for EWBG, the parameters of the SM cannot explain quantitatively the observed matter-antimatter asymmetry in the Universe. The electroweak phase transition in the SM is not strong enough, and an additional \(CP\)-violation source is required in the SM. Firstly, the electroweak phase transition is a strong first-order phase transition, if the following condition is satisfied \cite{Trodden:1998ym, Riotto:1998bt, Riotto:1999yt, Cohen:1993nk}, 
\begin{equation}\label{sphaleron_rate}
    \frac{\upsilon_c}{T_c} > 0.6 - 1.4,
\end{equation}
where \(T_c\) is the critical temperature of the electroweak phase transition and \(\upsilon_c\) is the vacuum expectation value (VEV) of the Higgs boson at this temperature. This condition expresses that EWBG can be achieved if the rate of sphaleron transitions is suppressed to avoid washing out the generated baryon asymmetry at the expanding bubble wall. It is important to mention that by convention the lower bound on this ratio is set to one, although it usually varies due to some theoretical uncertainties in the calculations \cite{Patel:2011th, Fuyuto:2014yia}.

Secondly, the EWBG relies on a \(CP\)-violating source that biases sphaleron interactions near the expanding bubble walls toward baryon production rather than anti-baryon production. In the real singlet extension, a \(CP\)-violating source could be introduced by an effective field theory with a dimension-six operator, considering the singlet as a dark matter candidate, as proposed in Ref. \cite{Cline:2012hg}. More specifically, in this effective field theory, the singlet is coupled with the top-quark mass, and the Yukawa interaction term in the SM Lagrangian is modified as
\begin{equation}\label{CP_source}
    y_t \bar{Q}_L H \left( 1 + \frac{\eta}{\Lambda^2} S^2 \right) t_R + \text{h.c.},
\end{equation}
where \(y_t\) is the top-quark Yukawa coupling, \(Q_L\) is the left-handed quark doublet, \(t_R\) is the right-handed top quark, and \(\Lambda\) is a new physics scale, which is much higher than 1 TeV. The complex phase, denoted as \(\eta\), is the \(CP\)-violating parameter\footnote{The \(CP\) violation can be maximized by taking \(\eta = i\) as this can be easily generalized to different phases, given that the baryon asymmetry depends linearly on the imaginary part.}. Then, the effective mass of the top quark is modified by the dimension-six operator in the Yukawa interaction term (\ref{CP_source}) and is written as
\begin{equation}\label{complextopquarkmass}
    m_t^2 (h, \phi) = \frac{y_t^2}{2} h^2 + \frac{y^2_t}{2} \frac{\eta^{*}}{\Lambda^2} h^2 \phi^2 + \frac{y^2_t}{2} \frac{\eta}{\Lambda^2} h^2 \phi^2 + \frac{y_t^2}{2} \frac{\eta^{*} \eta}{\Lambda^4} h^2 \phi^4.
\end{equation}
As a result, during the electroweak phase transition, the top quark has a complex mass that varies spatially along the profile of the bubble wall and reads
\begin{equation}
    m_t (z) = \frac{y_t}{\sqrt{2}} h(z) \left(1 + \eta \frac{\phi^2 (z)}{\Lambda^2}  \right),
\end{equation}
where \(z\) is the coordinate transverse to the bubble wall, in the limit that it has expanded large enough to be approximated as planar. The baryon asymmetry is computed in detail in Ref. \cite{Cline:2012hg}. In principle, a similar dimension-five operator to the dimension-six operator in (\ref{CP_source}) could provide a source of \(CP\) violation as shown in Ref. \cite{Espinosa:2011eu}, although it does not respect the \(\mathbb{Z}_2\) symmetry. Then, the two-loop contributions to the electric dipole moments of the electron and neutron should be studied to consider the relevant experimental constraints. However, this is not the case in this singlet extension with the dimension-six operators in (\ref{tree-level potential}) and (\(\ref{CP_source}\)) due to the forbidden Higgs-singlet mixing at zero temperature. Therefore, no additional experimental constraints are imposed on the complex phase owing to the forbidden Higgs-singlet mixing and the non-zero singlet VEV at high temperatures.

In addition, in the last years, the real singlet extension has been investigated for singlet masses \(m_S > m_H/2\) to avoid the experimental constraints by the invisible Higgs decay. We will examine these experimental constraints and the effect of the dimension-six operator (\ref{high-op}), which enhances the strength of the electroweak phase transition for low singlet masses. On the other hand, high energy physics has shifted its focus to gravitational wave experiments and astrophysical observations to probe EWBG and dark matter, imposing additional experimental constraints. EWBG could be probed by the stochastic gravitational wave background which is generated by a first-order phase transition \cite{Kosowsky:1991ua, Kamionkowski:1993fg, Huber:2008hg, Jinno:2016vai, Caprini:2007xq, Kosowsky:1992vn, Kosowsky:1992rz,Hindmarsh:2013xza, Hindmarsh:2015qta, Giblin:2013kea, Giblin:2014qia,Kahniashvili:2008pe, Kahniashvili:2009mf, Caprini:2009yp, Binetruy:2012ze}. Subsequently, gravitational wave experiments can be significant to explore Beyond the Standard Model (BSM) physics \cite{Huang:2016odd, Hashino:2022ghd, Apreda:2001us,Kusenko:2006rh,Chala:2018ari,Baldes:2016rqn,Noble:2007kk,Zhou:2020ojf, Weir:2017wfa, Hindmarsh:2020hop, Child:2012qg,LISACosmologyWorkingGroup:2022jok,Caprini:2015zlo,Huber:2015znp,Delaunay:2007wb,Grojean:2006bp,Katz:2014bha,Alves:2018jsw,Athron:2023xlk,Caprini:2019egz,NANOGrav:2023hvm,Ellis:2020awk,FitzAxen:2018vdt,Oikonomou:2023bah, Oikonomou:2024zhs, Oikonomou:2023qfz, Kakizaki:2015wua, Espinosa:2010hh, Bodeker:2009qy, Bartolo:2016ami} and are complementary probes with the collider searches, especially for the singlet extensions, \cite{Katz:2014bha, Alves:2018jsw, Noble:2007kk, Curtin:2014jma, Jain:2017sqm, Englert:2013tya, Craig:2013xia, Beniwal:2017eik}. 

The rest of the paper is organized as follows: In section II, we define the one-loop effective potential in the singlet extension of the SM with a dimension-six operator. In section III, the parameter space of our model is presented to discuss the various theoretical constraints, while section IV shows the experimental signatures to test EWBG in the real singlet extension. Finally, in section V, we investigate different scenarios for the electroweak phase transition in this singlet-extended SM for a wide range of singlet masses from \(0-550\) GeV and determine the parameter space to realize EWBG. The conclusions follow in the last section.

\section{Effective Potential}

\begin{figure}
\centering
\includegraphics[width=21pc]{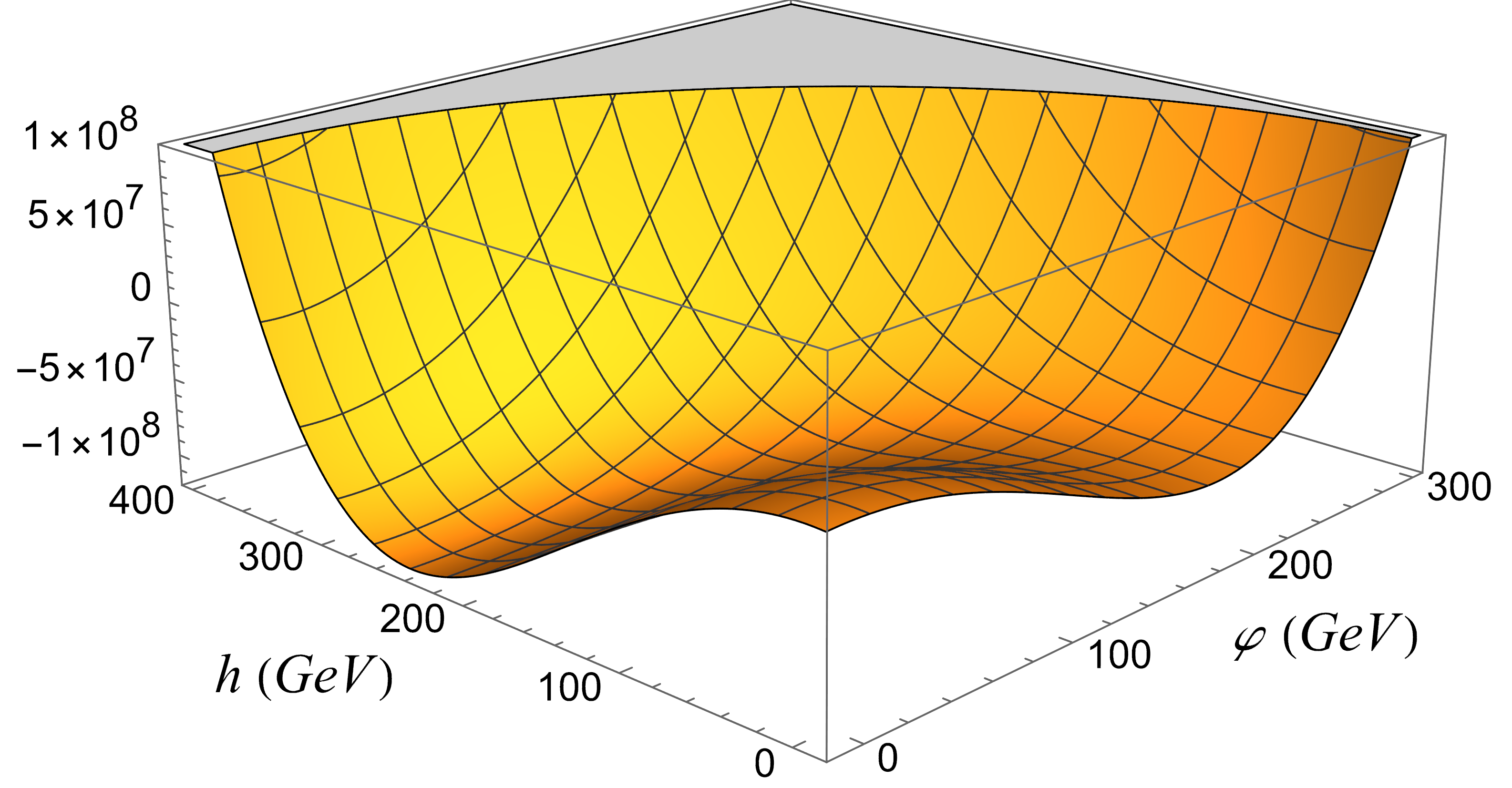}
\includegraphics[width=21pc]{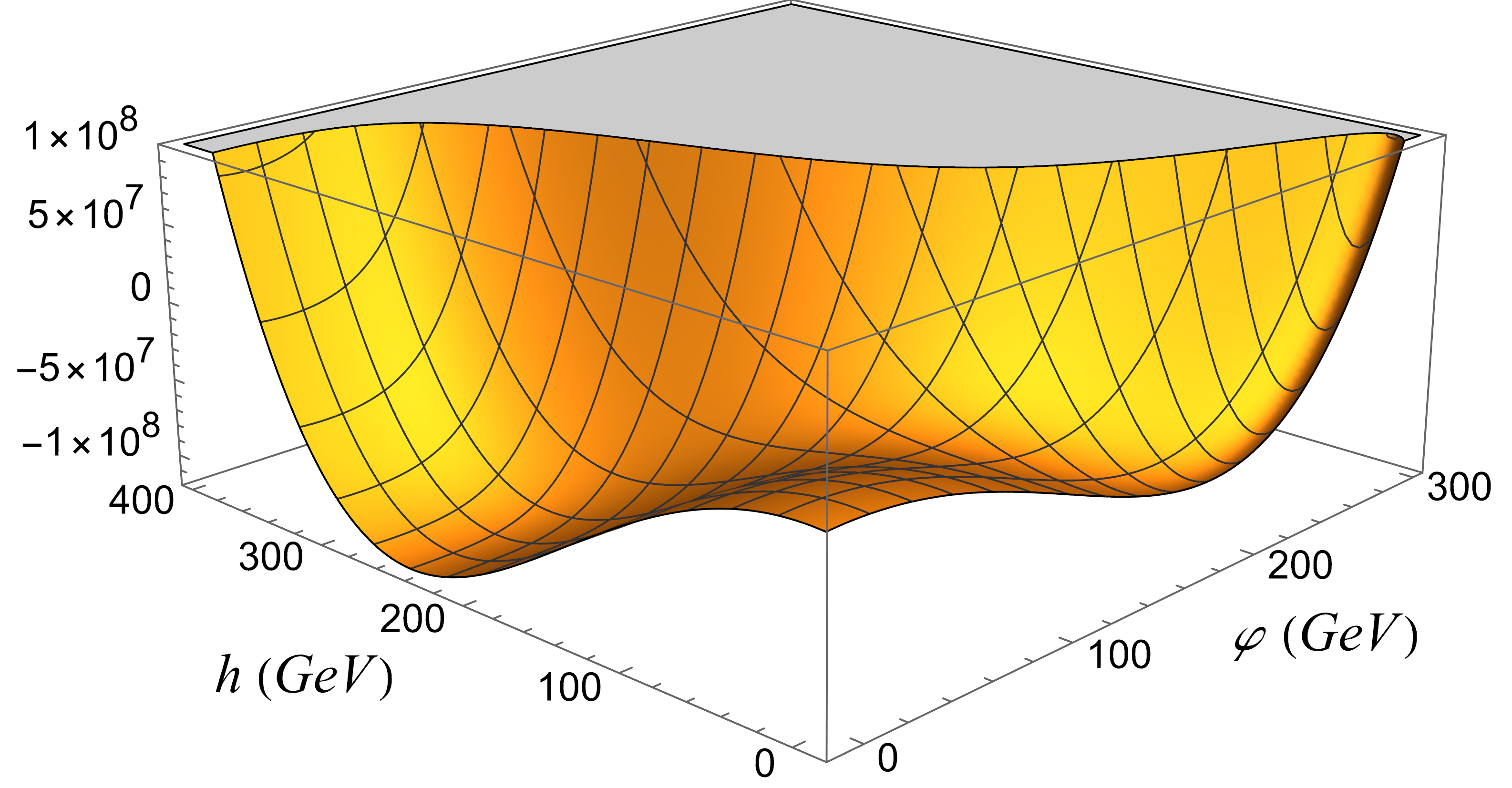}
\caption{The tree-level potential in the two-dimensional configuration space \(\left(h, \phi \right)\) for \(\lambda = 0\) (left) and \(\lambda/M^2 = 5 \times 10^{-5}\) GeV\(^{-2}\) (right), where the global minimum is the electroweak vacuum. It is assumed that \(m_S = 62.5\) GeV, \(\lambda_{HS} = 0.2\), and \(a =
0.1\).}\label{treelevelpotential}
\end{figure}

The dynamics of the phase transitions in the early Universe can be described by the effective potential, which consists of the tree-level contribution and higher loop contributions. In the SM, only the dominant contributions to the effective potential are considered that come from the gauge bosons, the top quark, and the Higgs and Goldstone bosons. The Higgs doublet can be parameterized as,
\begin{equation}\label{eq:1.2}
    H = \frac{1}{\sqrt{2}}\begin{pmatrix}
        \chi_1 + i \chi_2 \\
        \varphi + i \chi_3
        \end{pmatrix},
\end{equation}
where \(\varphi\) is the Higgs boson, and \(\chi_1, \chi_2\) and \(\chi_3\) are the three Goldstone bosons. The Higgs boson and the singlet scalar field can be shifted by a real constant background field, such as \(\varphi (x) \to h +  \varphi (x)\), to compute the effective potential using the background field method \cite{Jackiw:1974cv}. Then, the tree-level contribution to the effective potential is the tree-level potential expressed in terms of the background fields associated with the Higgs boson and the real singlet scalar field,
\begin{equation}\label{eq:1.3}
    V_0 (h,\phi) = - \frac{\mu^{2}_H}{2} h^2 + \frac{\lambda_H}{4} h^4 - \frac{\mu^2_S}{2}  \phi^2 + \frac{\lambda_S}{4} \phi^4 + \frac{\lambda_{HS}}{2} h^2 \phi^2 +  \frac{\lambda}{2M^2}h^2 \phi^4,
\end{equation}
which is shown in Fig. \ref{treelevelpotential} in the usual singlet extension (\(\lambda = 0\)) and the singlet extension with the higher-order operator (\(\lambda \ne 0\)). The VEV of the Higgs boson in the SM is
\begin{equation*}
    h = \upsilon =  \frac{\mu_H}{\sqrt{\lambda_H}}.
\end{equation*}
The effective masses can be derived by the SM Lagrangian and are given below in terms of the background fields,
\begin{equation}\label{effectivemasshiggs}
    m^2_h (h,\phi) = - \mu^2_H + 3\lambda_H h^2 + \lambda_{HS} \phi^2 + \frac{\lambda}{M^2}\phi^4, \quad m^2_{\chi} (h,\phi) = - \mu^2_H + \lambda_H h^2 + \lambda_{HS} \phi^2 +\frac{\lambda}{M^2}\phi^4,
\end{equation}
\begin{equation}\label{effectivemasssinglet}
    m^2_S (h,\phi) = - \mu^2_S + 3\lambda_S \phi^2 + \lambda_{HS} h^2 + \frac{6\lambda}{M^2}h^2 \phi^2,
\end{equation}
\begin{equation}\label{effectivemassW}
    m^2_W (h) = \frac{g^2}{4} h^2, \quad m^2_Z (h) = \frac{g^2 + g^{\prime 2}}{4} h^2, \quad m^2_t (h) = \frac{y^2_t}{2} h^2,
\end{equation}
where \(g\) and \(g^{\prime}\) are the \(SU(2)_L\) and \(U(1)_Y\) couplings, respectively\footnote{The numerical results are obtained by using the following masses at the vacuum state of the Universe \((h,\phi) = (\upsilon, 0)\): \(m_H = 125\) GeV, \(m_W = 80.4\) GeV, \(m_Z = 91.2\) GeV and \(m_t = 173\) GeV.}. 
In the next sections, the one-loop finite-temperature effective potential is computed in the Landau gauge to avoid the interaction between the Higgs boson and the ghost fields. The gauge-dependence of the effective potential will not be discussed here as it is out of the scope of this work and the reader is encouraged to see Refs. \cite{Yao:1973am, Nielsen:1975fs, Aitchison:1983ns, Espinosa:2016uaw, Ekstedt:2018ftj, Nielsen:2014spa, Kobes:1990dc}.

\subsection{Zero-Temperature Corrections}

The one-loop contribution to the effective potential at zero temperature is the Coleman-Weinberg (CW) potential \cite{Coleman:1973jx} and can be computed as the sum of all one-particle irreducible Feynman diagrams with zero external momenta and a single loop. In the \(\overline{\rm MS}\) renormalization scheme, the CW potential for each particle is written as,
\begin{equation}\label{eq:10}
    V^i_1 (h,\phi) = (-1)^{F_i} n_i \frac{m^4_{i}(h,\phi)}{64 \pi^2} \left[ \ln{\left( \frac{m^2_{i}(h,\phi)}{\mu^2_R}\right)} - C_i \right],
\end{equation}
where \(i = \{h, \chi, S, W, Z, t \}\) counts the fields in the extended SM, \(F_i = 1\) \((0)\) for fermions (bosons), \(n_i\) is the number of degrees of freedom of each field \(i\), \(\mu_R\) denotes the renormalization scale, and \(C_i = 3/2\) \((5/6)\) for scalars and fermions (gauge bosons). The number of degrees of freedom of each field is,
\begin{equation}\label{eq:11}
n_h = 1, \quad n_{\chi} = 3, \quad n_{S} = 1, \quad n_{W} = 6, \quad n_{Z} = 3, \quad n_{\gamma} = 2, \quad n_{t} = 12.
\end{equation}

\subsection{Finite-Temperature Corrections}
In 1972, Kirzhnits and Linde proposed that a gauge symmetry can be restored due to finite-temperature effects \cite{Kirzhnits:1972iw, Kirzhnits:1972ut, Kirzhnits:1976ts}. Namely, finite-temperature field theory is essential to derive the effective potential at finite temperatures and describe the \(SU(2) \times U(1)\) symmetry restoration in the early Universe. The one-loop effective potential at finite temperature consists of a temperature-dependent term and a temperature-independent term, which coincides with the effective potential at zero temperature shown in the previous subsection. As a result, the temperature-dependent component is briefly presented here and reads
\begin{equation}\label{T-potential}
V_T^i(h,\phi,T) = (-1)^{F_i} \frac{n_iT^4}{2 \pi^2} \int_{0}^{\infty} dx \, x^2 \ln \left[ 1 - (-1)^{F_i} \exp \left( - \sqrt{x^2 + \frac{m^2_{i}  (h,\phi)}{T^2} }\right) \right],
\end{equation}
where these integrals are the thermal functions defined as
\begin{equation}\label{thermalfunction}
J_{B/F} \left(y^2\right) = \int_{0}^{\infty} dx \, x^2 \ln \left[ 1 \mp \exp \left( - \sqrt{x^2 + y^2 }\right) \right],
\end{equation}
where the subscript \(B\) \((F)\) stands for bosons (fermions). The thermal functions can be computed numerically, but they also admit a high-temperature expansion \cite{Quiros:1999jp}, which will be applied in this work, and is given by,
\begin{equation}\label{bosonthermalfunction}
    J_B(y^2) =- \frac{\pi^4}{45} + \frac{\pi^2}{12} y^2 - \frac{\pi}{6} y^3 - \frac{1}{32} y^4 \log \left(\frac{y^2}{a_b} \right) + O (y^6),
\end{equation}
\begin{equation}\label{fermionthermalfunction}
    J_F(y^2) = \frac{7\pi^4}{360} - \frac{\pi^2}{24} y^2 - \frac{1}{32} y^4 \log \left(\frac{y^2}{a_f} \right) + O (y^6),
\end{equation}
where \(a_b = \pi^2 \exp{\left(3/2 - 2 \gamma_E \right)}\), \(a_f
= 16 \pi^2 \exp{\left(3/2 - 2 \gamma_E\right)}\), \(\zeta\)
denotes the Riemann \(\zeta\)-function, and \(\gamma_E\) is the
Euler-Mascheroni constant. The numerical evaluation of the thermal functions can prove that the high-temperature expansion up to the logarithmic term is accurate to better than \(5 \% \) for \(y \leq 2.2\) (\(y \leq 1.6\)) for bosons (fermions)  \cite{Laine:2016hma, Anderson:1991zb, Curtin:2016urg}.

Symmetry restoration at high temperatures implies that the one-loop perturbative analysis breaks down near the critical temperature \cite{Weinberg:1974hy}. More specifically, the dominant infrared contributions to the effective potential from higher-order diagrams should be included at all orders in the perturbative expansion \cite{Dolan:1973qd,Parwani:1991gq, Arnold:1992rz, Espinosa:1992gq, Kapusta:1984dx, Carrington:1991hz}. These higher-order diagrams are called ring diagrams, which are \(N\)-loop diagrams, where \(N-1\) of them are attached to the central one. In practice, the so-called thermal resummation is a method that includes higher-order corrections to the effective mass which is replaced by the thermal mass \cite{Parwani:1991gq}
\begin{equation}\label{def_thermal_mass}
    M^2 (h, T) = m^2 (h, \phi) + \Pi (h, T) \,,
\end{equation}
where \(m^2 ( h, \phi)\) is the effective mass-squared and \(\Pi(h, T)\) is the temperature-dependent self-energy \cite{Parwani:1991gq}. A method, developed by Arnold and Espinosa \cite{Arnold:1992rz}, resums only the zero Matsubara modes that cause the infrared divergence. In this scheme, the resummed effective potential is modified by introducing the following term,
\begin{equation}\label{rings}
    V^i_{ring} \left(m^2_i(h,\phi),T \right) = \frac{n_i T}{12\pi} \left[m^3_i(h,\phi) - \left( M^2_i (h, \phi, T)\right)^{3/2} \right],
\end{equation}
where \(i = \{h, \chi, S, W, Z, \gamma\}\) and \(\overline{n}_i = \{1, 3, 1, 2, 1, 1\}\) is the modified number of degrees of freedom which takes into account that only the longitudinal polarizations of the gauge bosons contribute to the temperature-dependent self-energy \cite{Arnold:1992rz, Carrington:1991hz}. In particular, the temperature-dependent self-energy for the scalar fields can be computed as,
\begin{equation}\label{T-higgs}
    \Pi_h (T) = \Pi_{\chi} (T) = \left(\frac{3g^2 }{16} + \frac{g^{\prime 2}}{16}  +\frac{y^2_t }{4} + \frac{\lambda_H}{2} + \frac{\lambda_{HS}}{12}\right) T^2,\quad
    \Pi_S (T) = \left( \frac{\lambda_S}{4} + \frac{\lambda_{HS}}{3} +  \frac{\lambda \upsilon^2}{2M^2} \right) T^2.
\end{equation}
The thermal corrections of the gauge bosons are treated differently because the transverse gauge fields have vanishing thermal corrections \cite{Carrington:1991hz}. Thus, the thermal gauge-boson mass matrix is computed and the corresponding eigenvalues are given by
\begin{equation}\label{Z-thermalmass}
    M^2_{Z_L/\gamma_L} (h, T) = \frac{1}{2} \left[ \frac{1}{4} \left(g^2 + g^{\prime 2}\right) h^2 + \frac{11}{6} \left(g^2 + g^{\prime 2} \right) T^2 \pm \sqrt{\left(g^2 - g^{\prime 2} \right)^2 \left( \frac{1}{4} h^2 + \frac{11}{6}  T^2 \right)^2  + \frac{g^2 g^{\prime 2}}{4} h^4 }\right],
\end{equation}
and the temperature-dependent self-energy of the longitudinal \(W\) bosons is
\begin{equation}\label{T-W}
    \Pi_{W_L} (T) = \frac{11}{6}g^2 T^2.
\end{equation}
In conclusion, the one-loop finite-temperature effective potential in the singlet-extended SM is written as,
\begin{equation}\label{generalfullpotential}
\begin{split}
    V_{\text{eff}} (h, \phi, T)  = & - \frac{\mu^{2}_H}{2} h^2 + \frac{\lambda_H}{4} h^4 - \frac{\mu^2_S}{2}  \phi^2 + \frac{\lambda_S}{4} \phi^4 + \frac{\lambda_{HS}}{2} h^2 \phi^2 + \lambda \frac{h^2 \phi^4}{2M^2} \\
    & + \sum_{i}  \frac{n_i m^4_{i}(h,\phi)}{64 \pi^2}\left[ \ln \left( \frac{m^2_{i}(h,\phi)}{\mu^2_R}\right) - C_i \right] -  \frac{n_t m^4_{t}(h)}{64 \pi^2}\left[ \ln \left( \frac{m^2_{t}(h)}{\mu^2_R}\right) - C_t \right] \\
    & + \sum_{i} \frac{n_iT^4}{2 \pi^2} J_{B} \left(\frac{m^2_i (h,\phi)}{T^2}\right) -  \frac{n_tT^4}{2 \pi^2} J_{F} \left(\frac{m^2_t (h)}{T^2}\right) \\
    & + \sum_{i} \frac{\overline{n}_i T}{12\pi} \left[m^3_i(h,\phi) - \left(M^2_i(h,\phi, T) \right)^{3/2} \right],
\end{split}
\end{equation}
where \(i = \{h, \chi, S, W, Z, \gamma \}\) represents the bosons in the singlet-extended SM. Most of our results rely on the high-temperature expansion (\ref{bosonthermalfunction}) and (\ref{fermionthermalfunction}) and the validity of this approach is mainly verified by the condition on the ratio \(M_i/T\), as previously discussed. The critical temperature \(T_c\) and the Higgs VEV \(\upsilon_c\) are computed numerically using this effective potential.

In this section, we did not take into account the \(CP\)-violating source that was introduced earlier in the real singlet extension to explain baryogenesis. It is evident that the effective potential is modified in the presence of an additional higher-order operator since the top-quark effective mass is then given by Eq. (\ref{complextopquarkmass}) instead of Eq. (\ref{effectivemassW}). The leading-order contribution of this operator to the one-loop finite-temperature effective potential using the high-temperature expansion reads,
\begin{equation}\label{cor-EFT}
    \delta V (h, \phi) = \frac{T^2}{4} \frac{y_t^2}{2} h^2 \left(\frac{\phi}{\Lambda}  \right)^4.
\end{equation}
As a consequence, if one considers this contribution, the critical temperature does not change significantly in the high-temperature limit as this contribution is higher than quadratic order in the background fields. Additionally, this contribution does not have an impact on the Higgs VEV at the critical temperature since it is zero in either field direction, along with its first derivatives. The presence of this dimension-six operator primarily results in a thinner bubble wall as the height of the barrier at the critical temperature increases. Finally, if the ratio \(\phi/\Lambda\) is assumed to be small (\(\Lambda > 1\) TeV), the correction (\ref{cor-EFT}) does not contribute significantly to the effective potential. Therefore, the dynamics of the phase transitions in the early Universe are not influenced considerably by the correction (\ref{cor-EFT}), which will not be taken into account in the effective potential. A similar discussion is presented in Refs. \cite{Cline:2012hg, Vaskonen:2016yiu}.

In this work, we do not explore phase transitions induced by curvature corrections to the one-loop effective potential since the curvature effects are either very small or vanish in the electroweak phase transition below the reheating temperature. Non-minimal coupling terms of the scalar fields and the scalar curvature, such as \(R \varphi^2\), can be introduced into the action, and the effective potential in the curved spacetime can be computed, as discussed in Refs. \cite{Buchbinder:1985jxp, Markkanen:2018pdo}. However, the electroweak phase transition occurs in the radiation-dominated era of the Universe \(\left(R=0\right)\) and these curvature corrections are more relevant during the inflationary epoch.

\section{Physical Parameter Space}
In this singlet extension of the SM, the parameter space is constructed by the following parameters: \(\mu_S^2\), \(\lambda_S\), \(\lambda_{HS}\), \(\lambda/M\), while it is assumed later for the numerical results that the effective theory is active at the energy scale \(M = 15\) TeV. The parameter space will be divided later based on the sign of \(\mu_S^2\). In addition, the parameter space of the model is primarily constrained by the vacuum structure, the validity of the perturbation theory, and the experimental measurements for the invisible Higgs decay width. If the model describes a strong electroweak phase transition, the parameter space is further eliminated by the condition for a strong electroweak phase transition  (\ref{sphaleron_rate}).

First of all, the current vacuum state of the Universe is considered to correspond to the Higgs VEV at zero temperature (\(h = \upsilon\)) with broken \(SU(2)_L \times U(1)_Y\) symmetry, whereas the VEV of the singlet scalar field vanishes (\(\phi = 0\)) and the \(\mathbb{Z}_2\) symmetry is unbroken. Namely, at zero temperature, the singlet has a positive mass squared given by
\begin{equation}\label{singlet_mass}
    m^2_{S} = - \mu^2_S + \lambda_{HS} \upsilon^2 > 0. 
\end{equation} 
As a result, we require that the Higgs minimum in the effective potential at zero temperature is the global minimum, and if the parameter \(\mu_S^2\) is positive, the following constraint is imposed
\begin{equation}\label{global_min}
    V_0 (\upsilon, 0) < V_0 (0, \upsilon_s) \Rightarrow \lambda_S > \lambda_H \frac{\mu^4_S}{\mu^4_H},
\end{equation}
where \(\upsilon_s\) is the singlet VEV at zero temperature in the \(\phi\) direction. One concludes from Eq. (\ref{singlet_mass}) that if \(\mu_S^2\) is positive, the Higgs-singlet coupling is always positive. Furthermore, it is imposed that \(\lambda_S > 0\) and \(\lambda > 0\) since the tree-level potential should be bounded from below. Subsequently, in view of Eqs. (\ref{global_min}) and (\ref{singlet_mass}), one can derive the minimum value of the singlet self-coupling,
\begin{equation}\label{lambda_min}
    \lambda^{\text{min}}_S = \frac{\lambda_H}{\mu^4_H} \left(m^2_S - \lambda_{HS} \upsilon^2 \right)^2
\end{equation}
so that the singlet self-coupling is written as
\begin{equation}\label{lambda_definition}
    \lambda_S = \lambda^{\text{min}}_S + a,
\end{equation}
where \(a\) is a positive parameter, which usually takes the value \(a = 0.1\) for singlet masses \(m_S \geq m_H/2\) \cite{Curtin:2014jma, Chiang:2018gsn, Jain:2017sqm}. Hence, we could equivalently express the parameter space in terms of the singlet mass, the Higgs-singlet interaction coupling, and the Wilson coefficient for a given parameter \(a\). However, if \(\mu_S^2>0\), the one-loop perturbative analysis is not reliable for \(\lambda_S > 8\), which implies that \(\lambda_{HS} \gtrsim 6\) with the singlet mass ranging from \(0 - 550\) GeV \cite{Curtin:2014jma}. This result is based on approximate calculations to show the regions of the parameter space, where we cannot trust the one-loop perturbative analysis. Although the Renormalization Group evolution of this model is important for a more detailed analysis, its results do not change significantly the parameter space. In this study, it is also important to require that the Wilson coefficient is \(\lambda/M^2 < 10^{-4}\) GeV\(^{-2}\) as the effective field theory is not a strongly coupled theory.

Now avoiding negative runaway directions in the tree-level potential leads to an additional constraint,
\begin{equation}\label{negmucon}
   \lambda_S > \frac{\lambda_{HS}^2}{\lambda_H} \quad \Rightarrow \quad
    \lambda_{HS} > - \sqrt{\lambda_H \lambda_S}.
\end{equation}
This constraint is important in the case of negative \(\lambda_{HS}\) which implies that the parameter \(\mu_S^2\) is negative. Thus, if we follow a similar approach to the previous scenario, for a given negative \(\lambda_{HS}\), the singlet's quartic coupling can be expressed as
\begin{equation}
    \lambda_{S}^{'} = \frac{\lambda_{HS}^{2}}{\lambda_{H}} + a.
\end{equation}
Nevertheless, if \(\mu_S^2 < 0\), the one-loop approximation is not valid for large values of the Higgs-singlet coupling, such as \(\lambda_{HS} \gtrsim 4\) for singlet masses up to \(550\) GeV. Accordingly, if we require that \(\lambda^{\text{min}}_S < 8\), the constraint (\ref{negmucon}) implies that \(\lambda_{HS} \gtrsim -1.0 \) which is a condition for a valid one-loop perturbative analysis.

\section{Experimental Signatures}
\begin{figure}[h!]
\centering
\includegraphics[width=20.8pc]{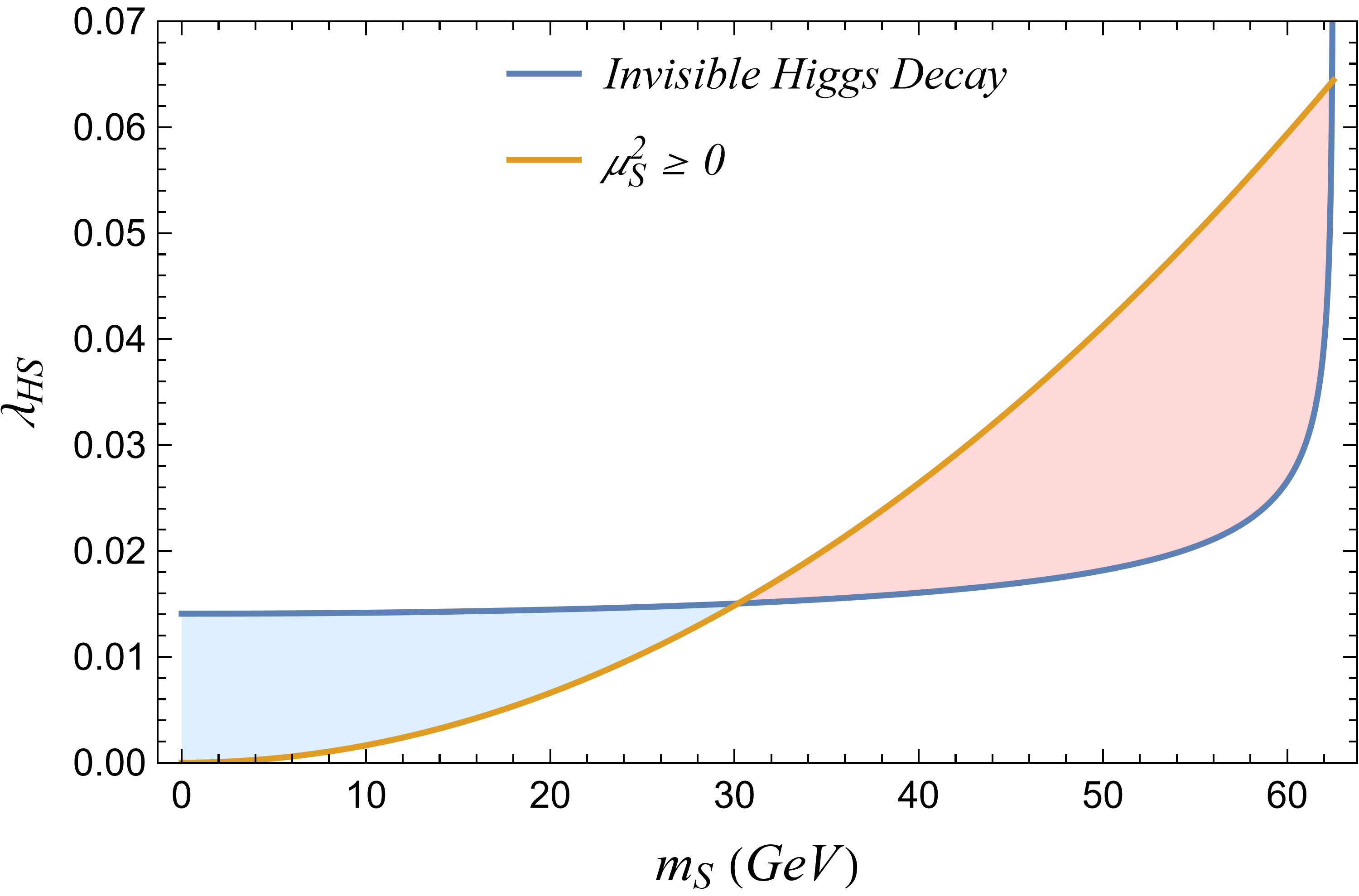}
\includegraphics[width=20.8pc]{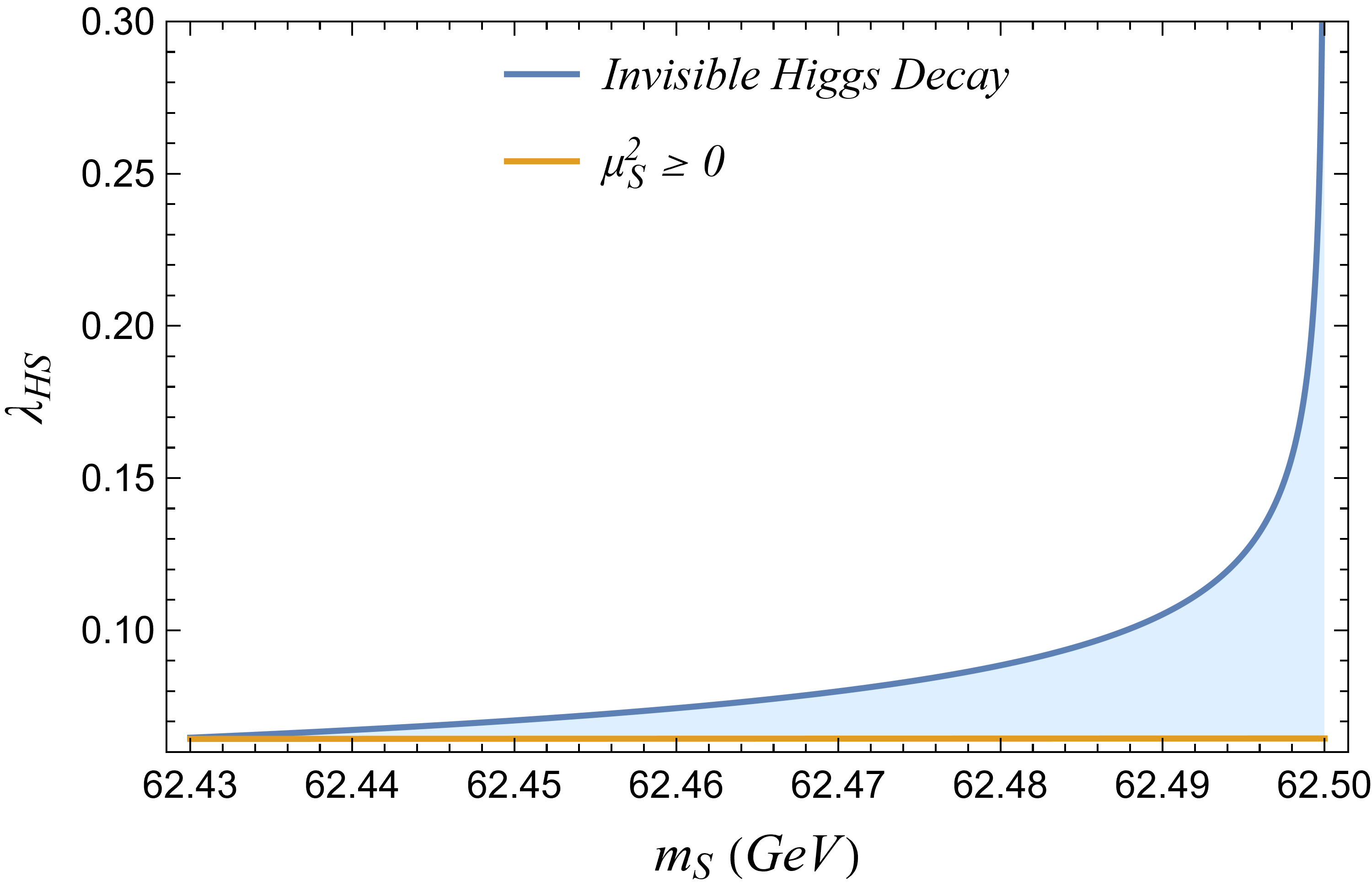}
\caption{The parameter space (left) for \(m_S < 62.5 \text{ GeV}\) is the blue shaded region, which only satisfies the constraint (\ref{condition_coupling}). The higher mass region (right) is also presented in detail to show the wide range of values of \(\lambda_{HS}\) with \(m_S > 62.43\) GeV.} \label{BR_0.19}
\end{figure}

Now we shall present the experimental signatures for the EWBG described by the real singlet extension and analyze further the constraints imposed by the current collider searches for the invisible decays of the Higgs boson. The singlet extensions of the SM are mainly studied for \(m_S > m_H/2\) to rule out the experimental constraints by the invisible Higgs decays. Namely, current and future experiments can impose strong constraints on the parameter space of this extension, if \(m_S < m_H/2\) which implies that the decay \(H \to S S\) becomes kinematically allowed. We will illustrate how the results from collider experiments for the branching ratio of the Higgs to the invisible sector can be translated into constraints for the parameter space of the singlet extension. The lower bound of the branching ratio of the Higgs to the invisible sector has been measured by the ATLAS \cite{ATLAS:2020kdi} and CMS \cite{CMS:2018yfx} experiments to be \(BR_{\text{inv}} < 0.11 - 0.19\) at \(95 \% \) CL \cite{ParticleDataGroup:2022pth}. The most recent combination of searches for invisible Higgs decays shows that \(BR_{\text{inv}} < 0.107\) at \(95\%\) CL \cite{ATLAS:2023tkt}. In general, the invisible decay width of the Higgs boson is computed as 
\begin{equation}\label{decay_width}
    \Gamma (H \to S S) = \frac{\lambda^2_{HS} \upsilon^2}{32 \pi m_H} \sqrt{1 - \frac{4 m^2_S}{m^2_H}}.
\end{equation}
If we denote the upper bound on the invisible decay width of the Higgs boson as \(\Gamma_{\text{m}}\), this upper bound results in an upper bound on the Higgs-singlet coupling \(\lambda_{HS}\),
\begin{equation}\label{special_condition_coupling}
    \lambda_{HS} < \sqrt{ \frac{32 \pi m_H}{\upsilon^2}\left({1 - \frac{4 m^2_S}{m^2_H}} \right)^{-1/2}\Gamma_{\text{m}}}.
\end{equation}
Additionally, if we assume that \(\mu^2_S \geq 0\), the constraint (\ref{special_condition_coupling}) leads to
\begin{equation}\label{condition_coupling}
    \frac{m_S^2}{\upsilon^2} \leq \lambda_{HS} < \sqrt{\frac{32 \pi m_H}{\upsilon^2}\left({1 - \frac{4 m^2_S}{m^2_H}} \right)^{-1/2} \Gamma_{\text{m}} },
\end{equation}
which does not depend on the singlet's self-coupling and the Wilson coefficient. For instance, if the branching ratio of the Higgs boson to the invisible particles is set to \(BR_{\text{inv}} < 0.19\) and the Higgs decay width to visible channels is \(\Gamma_{\text{vis}} = 4.07\) MeV, then the upper bound on the invisible decay width of the Higgs boson is,
\begin{equation}\label{decay_width_0.19}
    \Gamma (H \to SS) < 0.955 \text{ MeV}.
\end{equation}
As a result, the parameter space, which is allowed by the condition (\ref{condition_coupling}), is illustrated in Fig. \ref{BR_0.19}. The lower mass region (\(m_S < 30.19\) GeV) has the upper bound of \(\lambda_{HS} = 0.014\), whereas the higher
mass region (\(m_S > 62.43\) GeV) has a wider range of values for
\(\lambda_{HS}\).
In the next section, it will be demonstrated that low values of the Higgs-singlet coupling result in a smooth electroweak phase transition. Thus, the parameter space in Fig. \ref{BR_0.19} is further restricted, considering the condition (\ref{sphaleron_rate}) for a strong phase transition and the lower bound on the Higgs-singlet coupling may be higher than the one in Fig. \ref{BR_0.19}.

The direct production of singlet states can be probed by colliders, although it has been studied that this direct detection requires a \(100\) TeV collider \cite{Curtin:2014jma}, especially for probing the parameter space for the strong two-step phase transition in the high-mass region (\(m_S>m_H/2\)). In contrast, indirect collider signatures are by far more promising since colliders can probe the modifications to the triple Higgs coupling and the \(Zh\) cross section that arise from the real singlet scalar field \cite{Katz:2014bha, Alves:2018jsw, Noble:2007kk, Curtin:2014jma, Jain:2017sqm, Englert:2013tya, Craig:2013xia, Beniwal:2017eik}. The modification of the triple Higgs coupling is obtained by the third derivative of the one-loop zero-temperature effective potential,
\begin{equation}
    \lambda_3 = \frac{1}{6} \left. \frac{\partial^3 V_{\text{eff}} (h, \phi = 0, T = 0)}{\partial h^3} \right|_{h = \upsilon} \simeq \frac{m^2_H}{2 \upsilon^2} + \frac{\lambda_{HS}^3 \upsilon^3}{24 \pi^2 m_S^2},
\end{equation}
where the first term comes from the tree-level potential (\ref{tree-level potential}) and the singlet one-loop contribution results in the second term, omitting the logarithmic term. This coupling is measurable at the HL-LHC during double Higgs production events, but the extremely low cross section poses significant challenges for this measurement \cite{Beniwal:2017eik}. Moreover, the singlet can influence the Higgs couplings by introducing a slight correction to the Higgs wave function renormalization, resulting in modifications to all Higgs couplings by a potentially detectable amount. More specifically, precision measurements of the \(Zh\) production cross section at lepton colliders could offer an additional method for the indirect detection of the singlet \cite{Englert:2013tya, Craig:2013xia}. However, the measurement of the triple Higgs coupling is a better probe than the \(Zh\) production \cite{Beniwal:2017eik}, although they complement each other for the indirect detection of the singlet \cite{Curtin:2014jma}.

Additionally, the tree-level potential (\ref{tree-level potential}) possesses a \(\mathbb{Z}_2\) discrete symmetry, enabling the singlet to serve as a dark matter candidate \cite{Alanne:2014bra, Cline:2013gha, Ghorbani:2018yfr, Ghorbani:2020xqv, Cline:2012hg, GAMBIT:2017gge,Athron:2018ipf,Feng:2014vea, Beniwal:2017eik, Profumo:2007wc, Gonderinger:2009jp}. As a result, the unbroken \(\mathbb{Z}_2\) symmetry at zero temperature guarantees the stability of the dark matter particle and prevents any mixing between the Higgs field and the singlet scalar field. Therefore, further constraints from electroweak precision tests and Higgs coupling modifications are excluded as experimental signatures. Furthermore, considering the singlet as a dark matter candidate is not consistent with a strong electroweak phase transition in most regions of the parameter space of the real singlet extension as the parameter space is highly constrained by the direct dark matter searches carried out by LUX, XENON1T, and XENONnT experiments \cite{Cline:2013gha, Cline:2012hg, GAMBIT:2017gge, Athron:2018ipf, Feng:2014vea, Beniwal:2017eik}. However, around the Higgs resonance region (\(m_S = m_H/2\)), the singlet can account for a proportion of the total observed dark matter density \cite{GAMBIT:2017gge, Athron:2018ipf, Feng:2014vea}.

Last but not least, gravitational wave experiments could test EWBG described by BSM physics \cite{Huang:2016odd, Hashino:2022ghd, Apreda:2001us,Kusenko:2006rh,Chala:2018ari,Baldes:2016rqn,Noble:2007kk,Zhou:2020ojf, Weir:2017wfa, Hindmarsh:2020hop, Child:2012qg,LISACosmologyWorkingGroup:2022jok,Caprini:2015zlo,Huber:2015znp,Delaunay:2007wb,Grojean:2006bp,Katz:2014bha,Alves:2018jsw,Athron:2023xlk,Caprini:2019egz,NANOGrav:2023hvm,Ellis:2020awk,FitzAxen:2018vdt,Oikonomou:2023bah, Oikonomou:2024zhs, Oikonomou:2023qfz, Kakizaki:2015wua, Beniwal:2017eik, Vaskonen:2016yiu, Espinosa:2010hh, Bodeker:2009qy, Bartolo:2016ami}. A first-order phase transition is a very violent process, which involves enormous transfers of energy, and a stochastic gravitational wave background is generated during a first-order phase transition \cite{Kosowsky:1991ua, Kamionkowski:1993fg, Huber:2008hg, Jinno:2016vai, Caprini:2007xq, Kosowsky:1992vn, Kosowsky:1992rz,Hindmarsh:2013xza, Hindmarsh:2015qta, Giblin:2013kea, Giblin:2014qia,Kahniashvili:2008pe, Kahniashvili:2009mf, Caprini:2009yp, Binetruy:2012ze}. In fact, in recent years, a number of theoretical studies have focused on computing the gravitational wave background from the electroweak phase transition in the real singlet extension and predicting its detection in future gravitational wave experiments, such as the Laser Interferometer Space Antenna (LISA) and the Big Bang Observer (BBO) interferometers \cite{Caprini:2015zlo, Caprini:2019egz, Ellis:2020awk, Athron:2023xlk, Alves:2018jsw, Beniwal:2017eik, Vaskonen:2016yiu, Kakizaki:2015wua, Bartolo:2016ami}. The gravitational wave signals generated by first-order phase transitions are further discussed in Appendix A, while the stochastic gravitational wave background from first-order phase transitions in our singlet-extended SM is left for future studies. 

\section{Electroweak Phase Transition}
In this section, we demonstrate the behavior of the finite-temperature effective potential in the real singlet extension of the SM and discuss further the parameter space that complies with the experimental and theoretical constraints and realizes a strong electroweak phase transition.

In the early Universe, the electroweak symmetry is restored due to thermal fluctuations. As the temperature of the Universe drops below a critical temperature, the Higgs VEV acquires a non-zero value, which determines the vacuum state of the Universe, and the electroweak symmetry is spontaneously broken. In particular, at high temperatures, the effective potential in the SM is minimized for a vanishing Higgs VEV, but as the temperature decreases, a non-zero local minimum appears, which is initially separated by a potential barrier and eventually becomes the global minimum. As a consequence, a first-order phase transition occurs from the symmetric phase with \(\langle H\rangle = 0\) to the phase of the true vacuum with \(\langle H \rangle \ne 0\). A first-order phase transition from the false to the true vacuum proceeds via bubble nucleation, while bubbles of the broken phase are formed and expand throughout the Universe, converting the false vacuum into the true one.

\begin{figure}[h!]
\centering
\includegraphics[width=14pc]{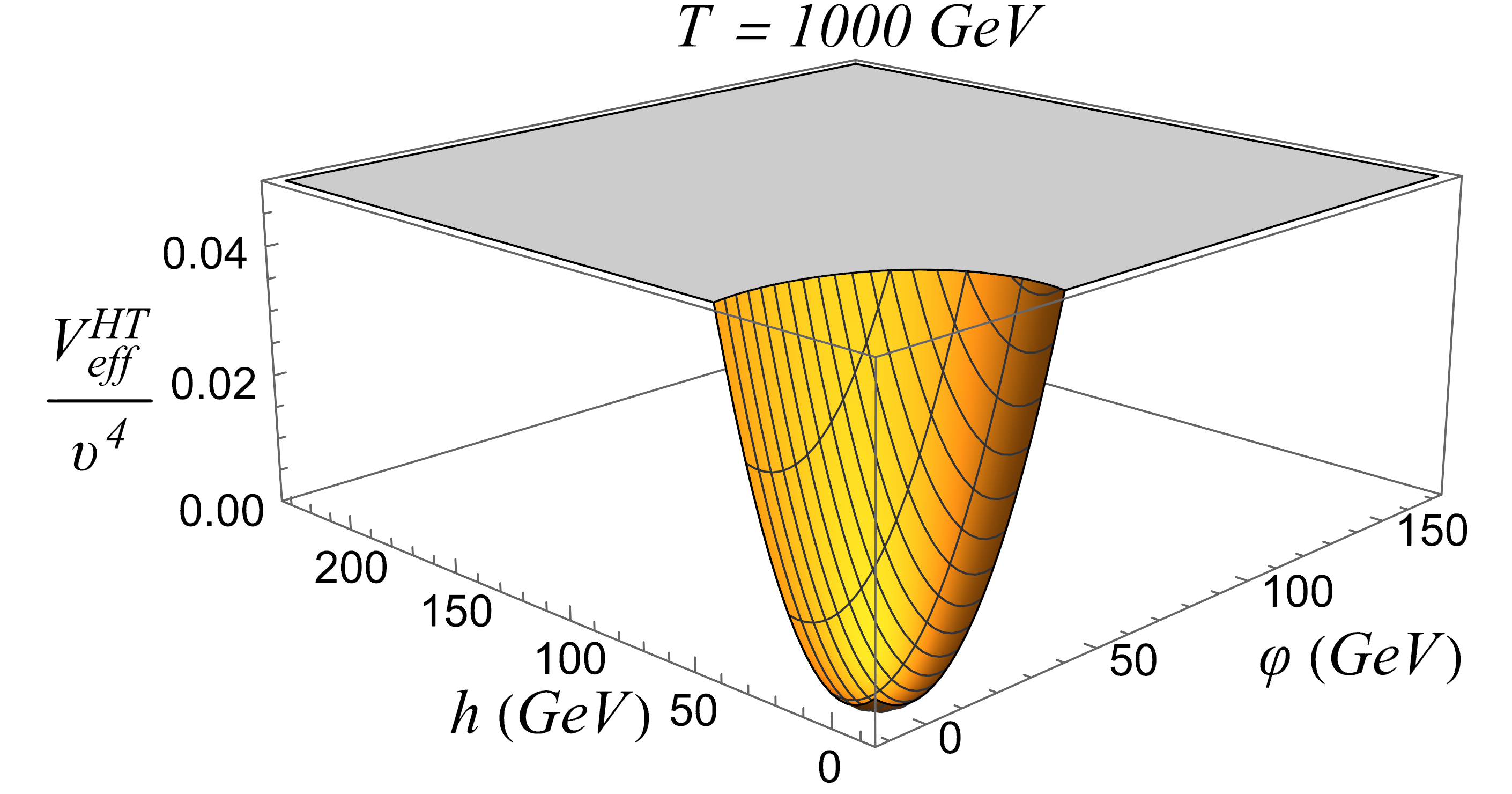}
\includegraphics[width=14pc]{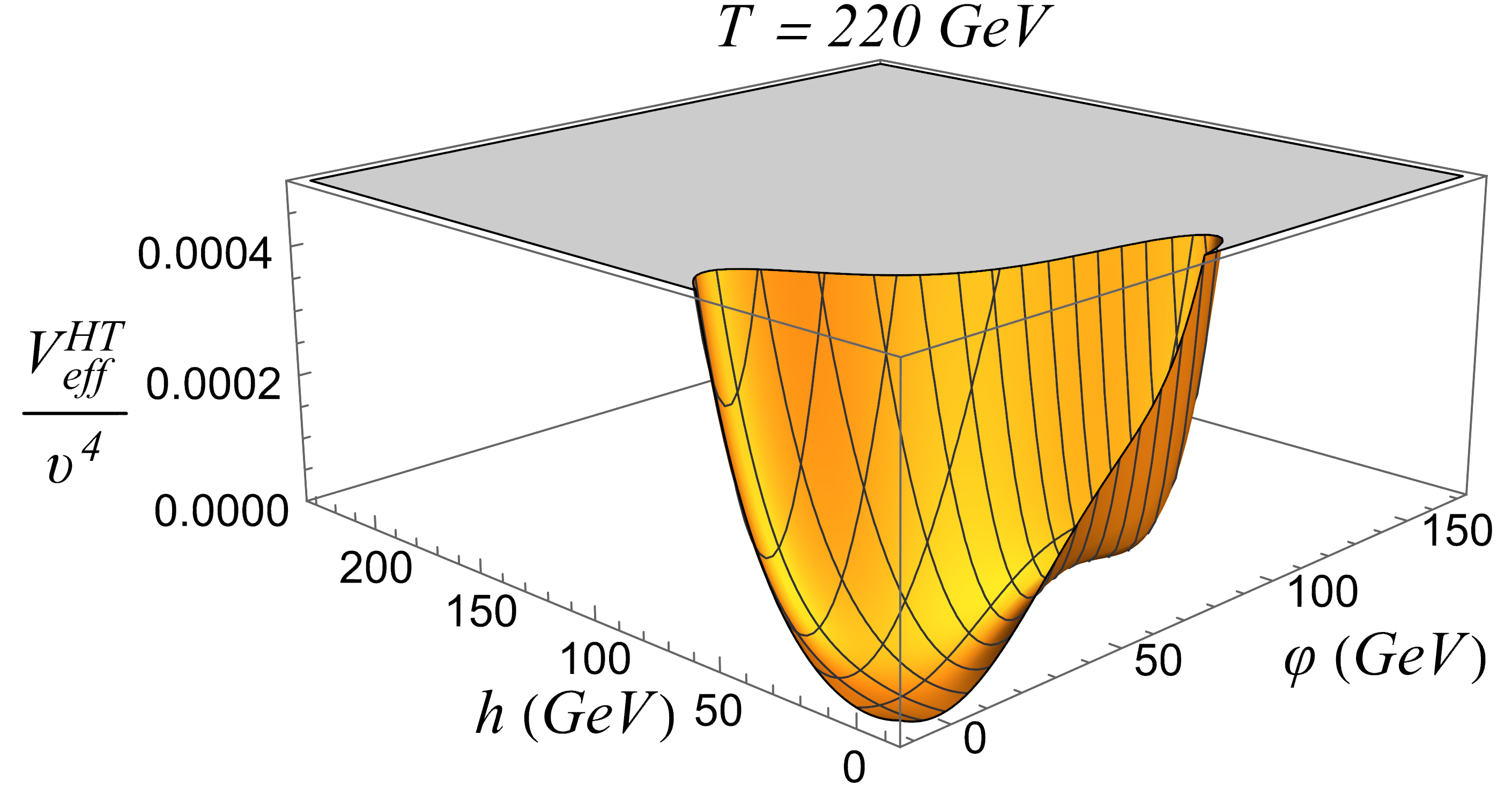}
\includegraphics[width=14pc]{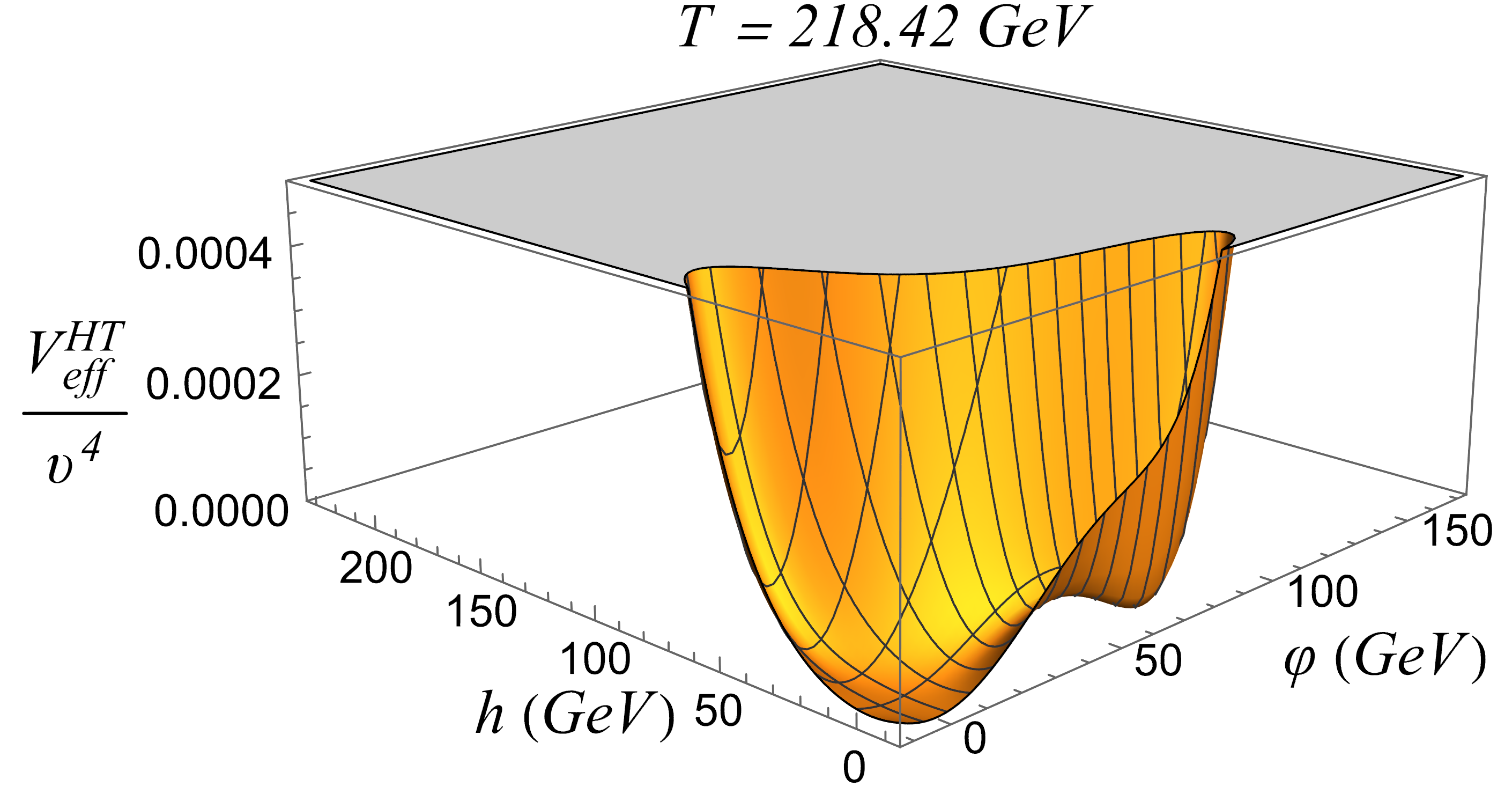}
\includegraphics[width=14pc]{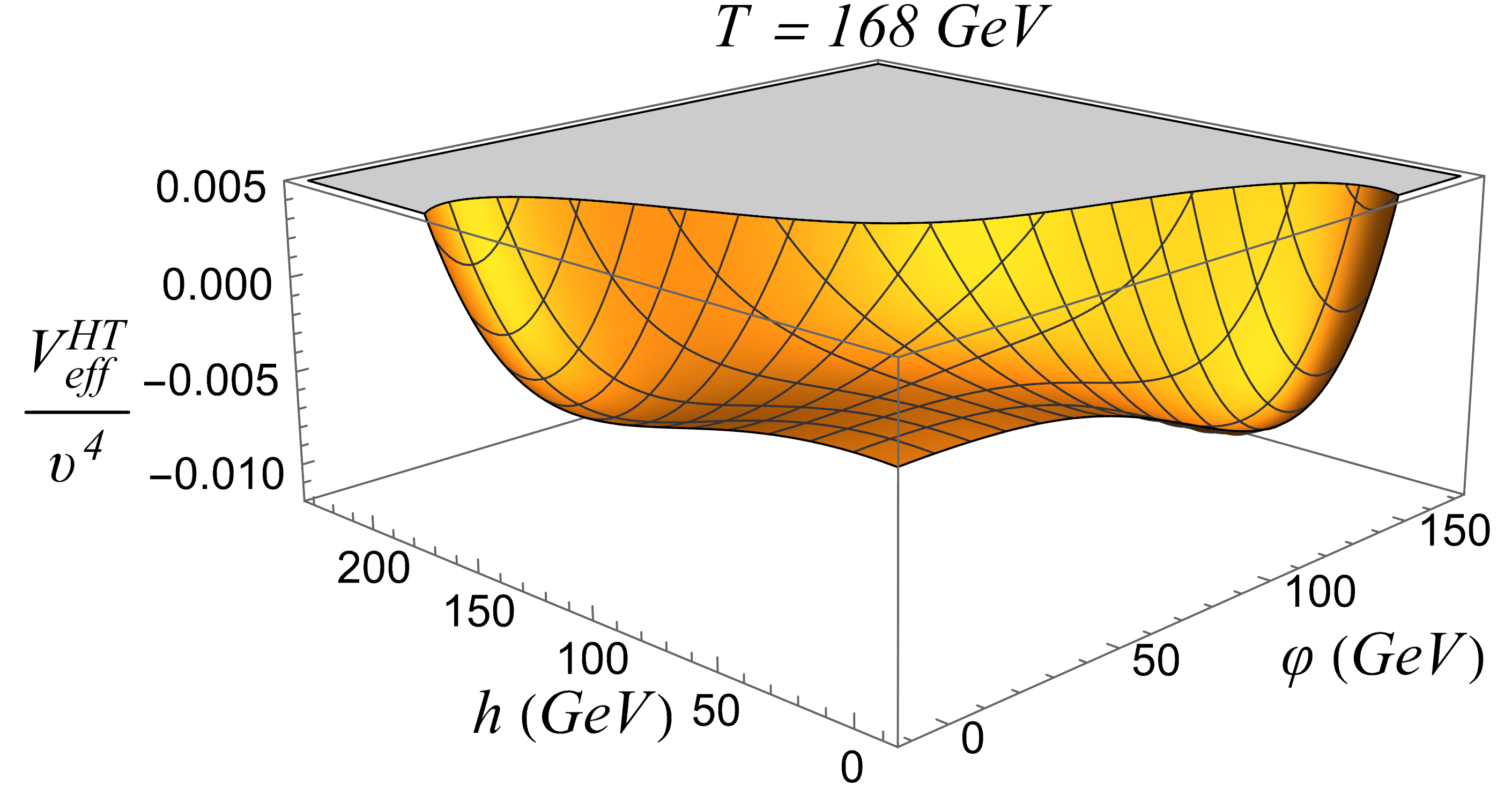}
\includegraphics[width=14pc]{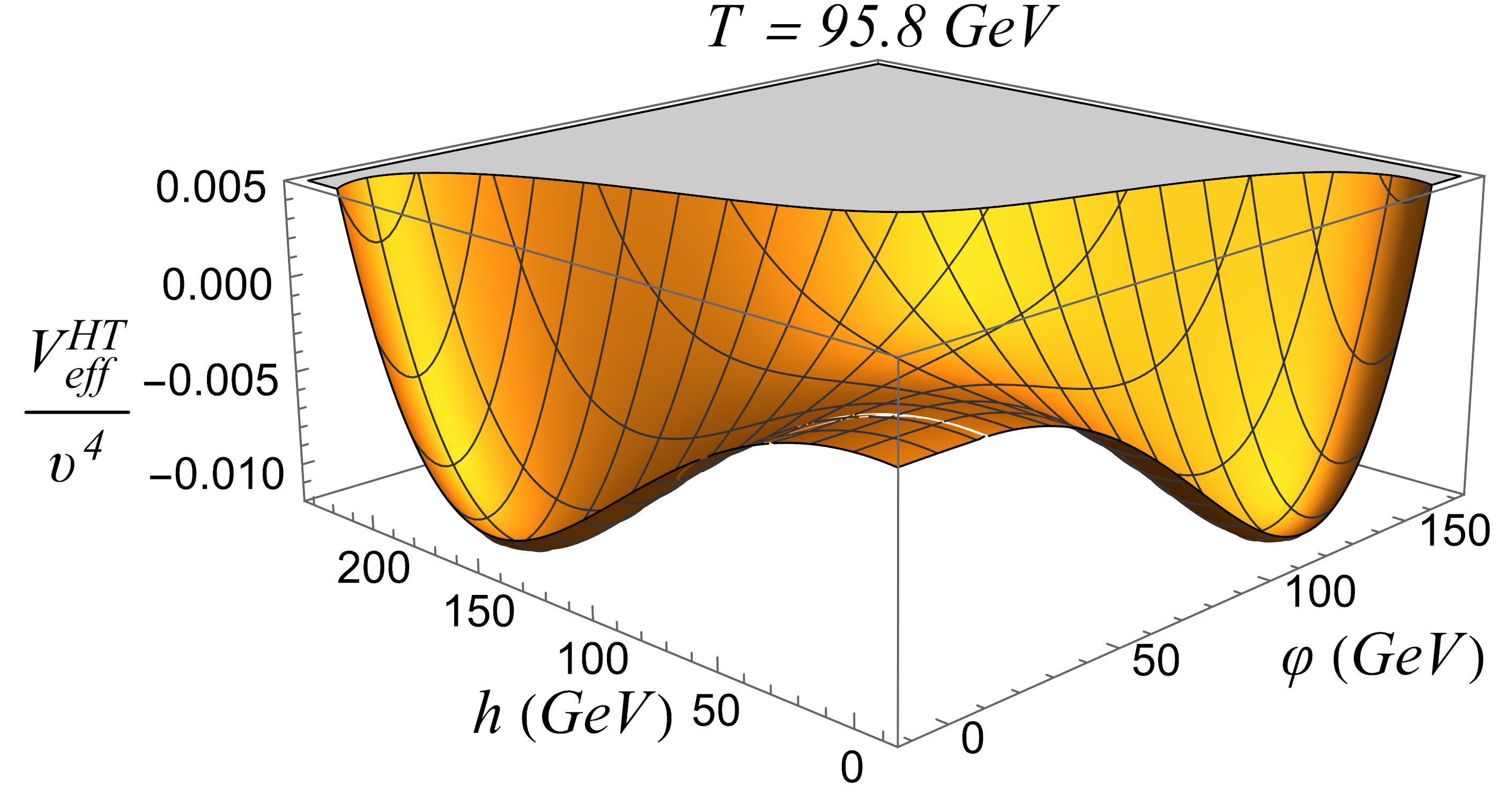}
\includegraphics[width=14pc]{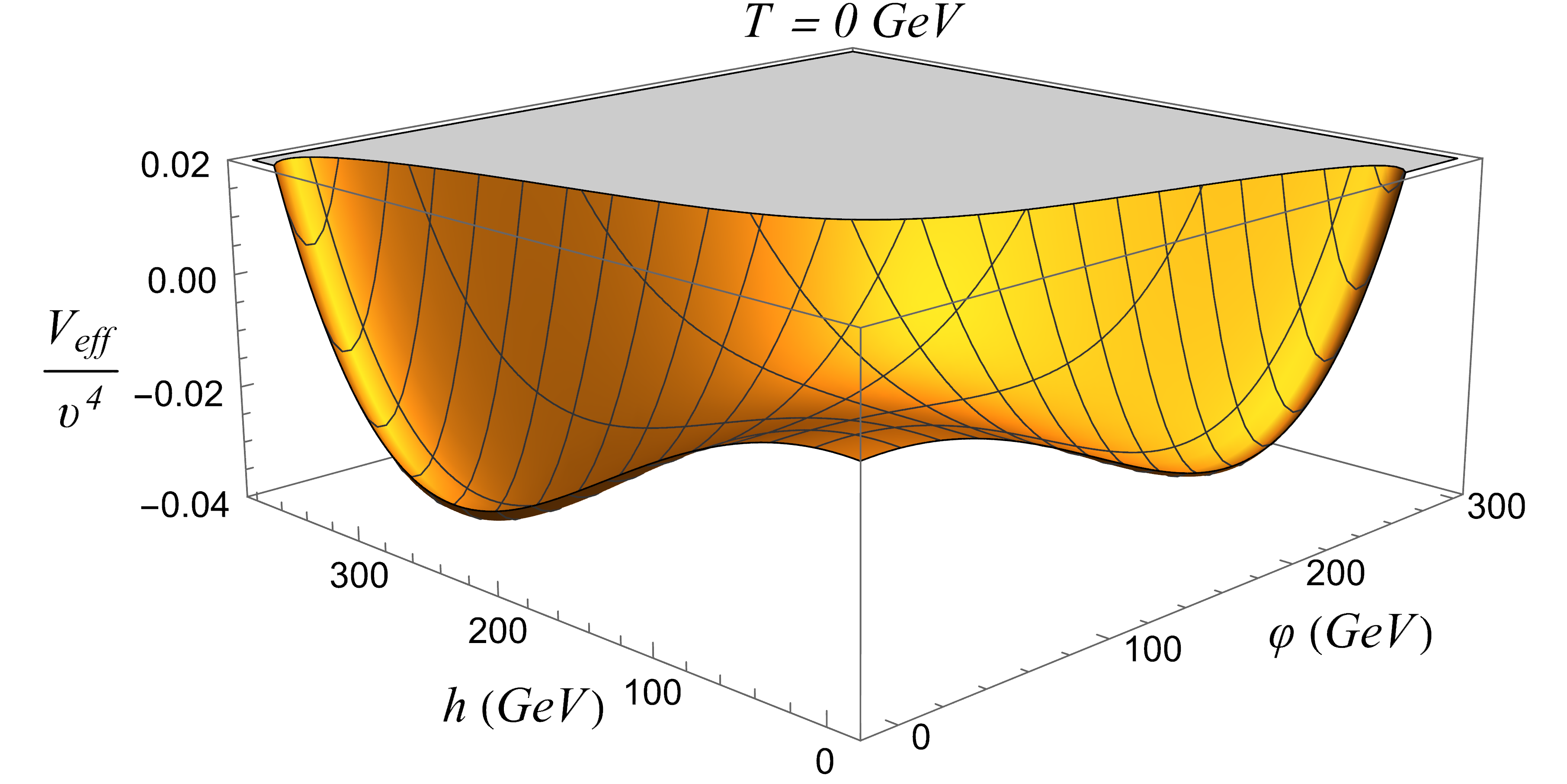}
\caption{The effective potential as the Universe cools down. In this
example, the singlet's phase transition is first-order using a
point of the parameter space with \(m_S = 500\) GeV,
\(\lambda_{HS} = 4.3\), \(\lambda/M^2 \simeq 2 \times 10^{-5}\)
GeV\(^{-2}\), and \(a = 0.1\).} \label{T-potential_2}
\end{figure}

\begin{figure}[h!]
\centering
\includegraphics[width=14pc]{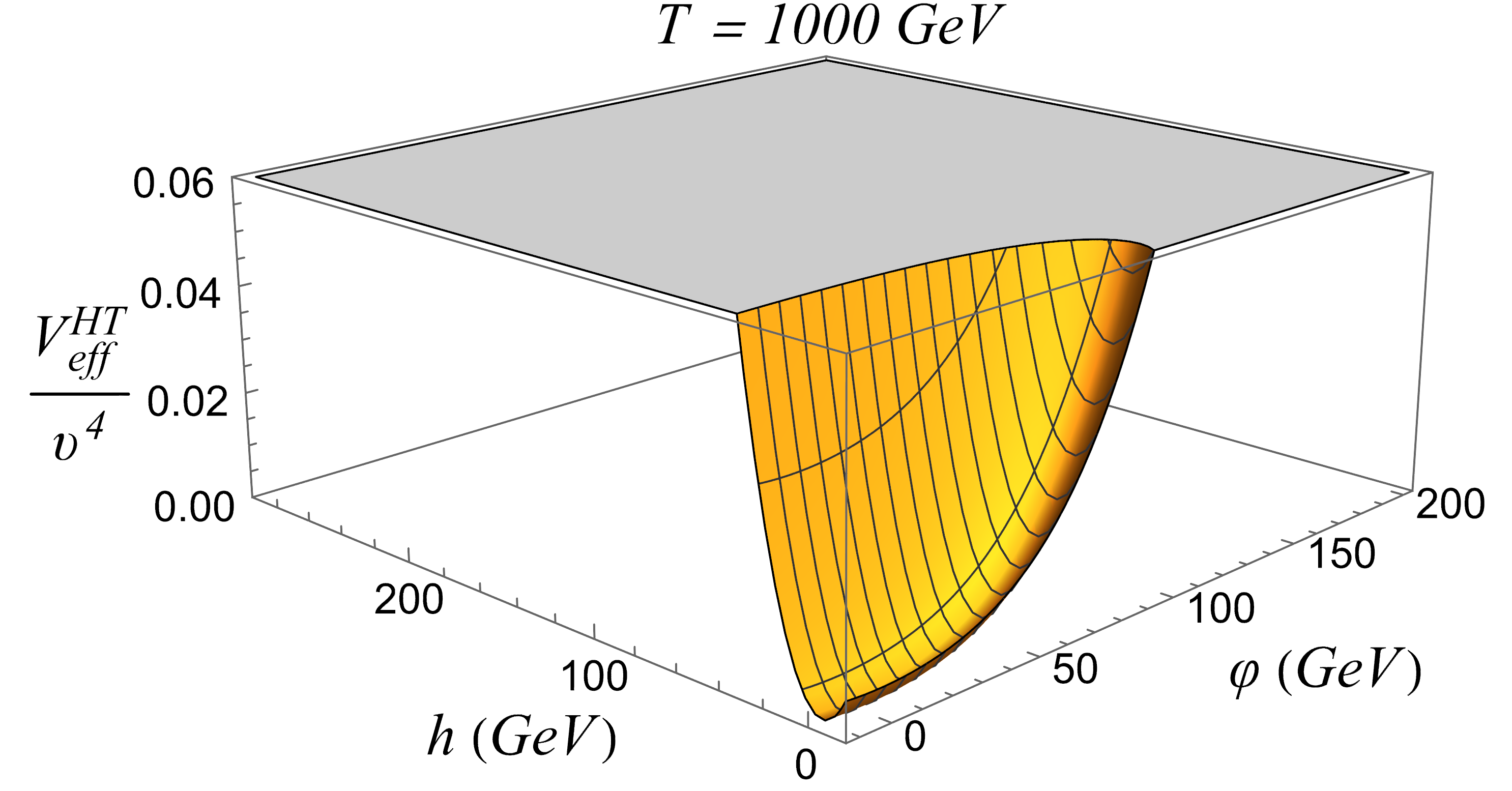}
\includegraphics[width=14pc]{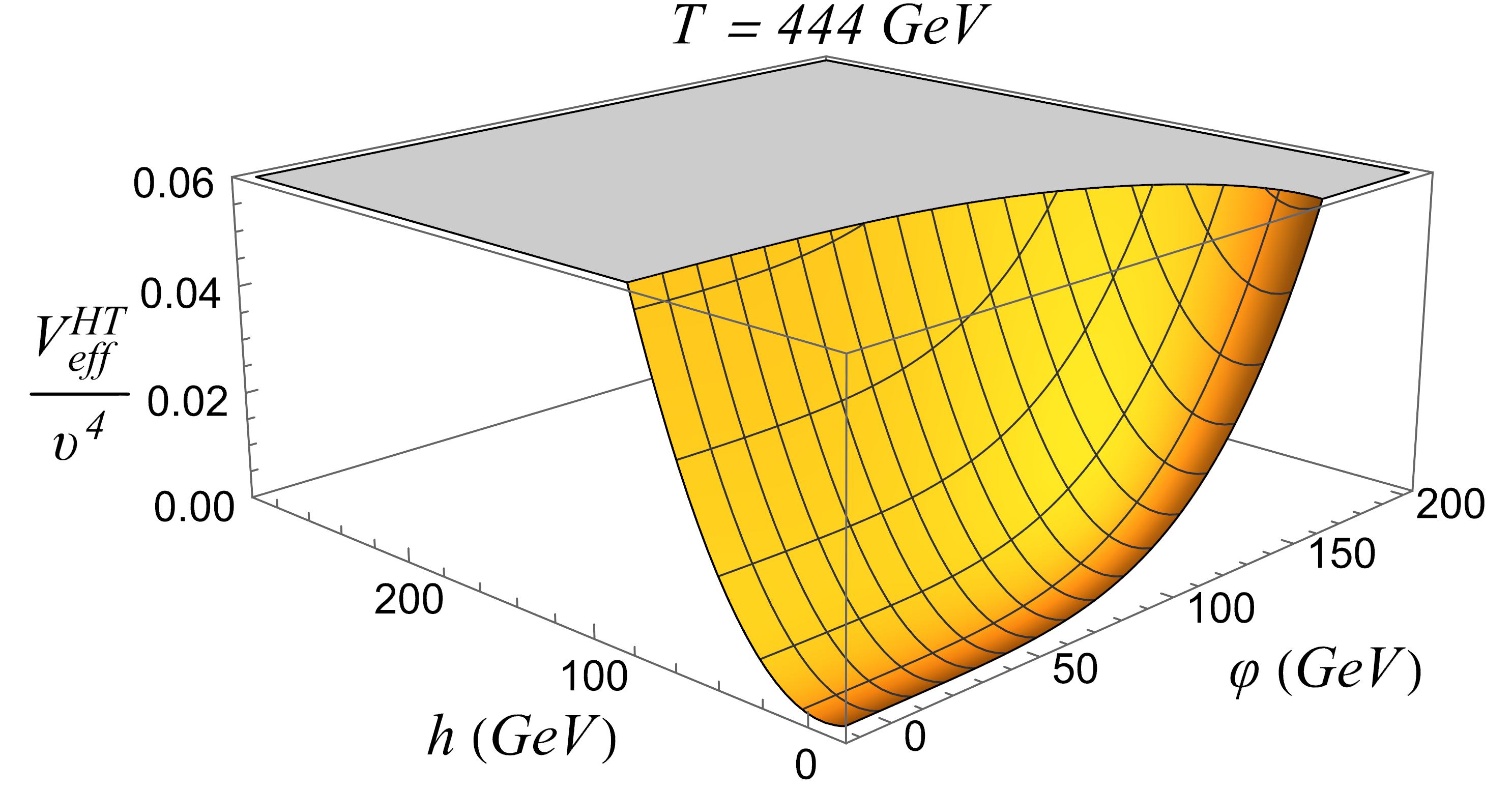}
\includegraphics[width=14pc]{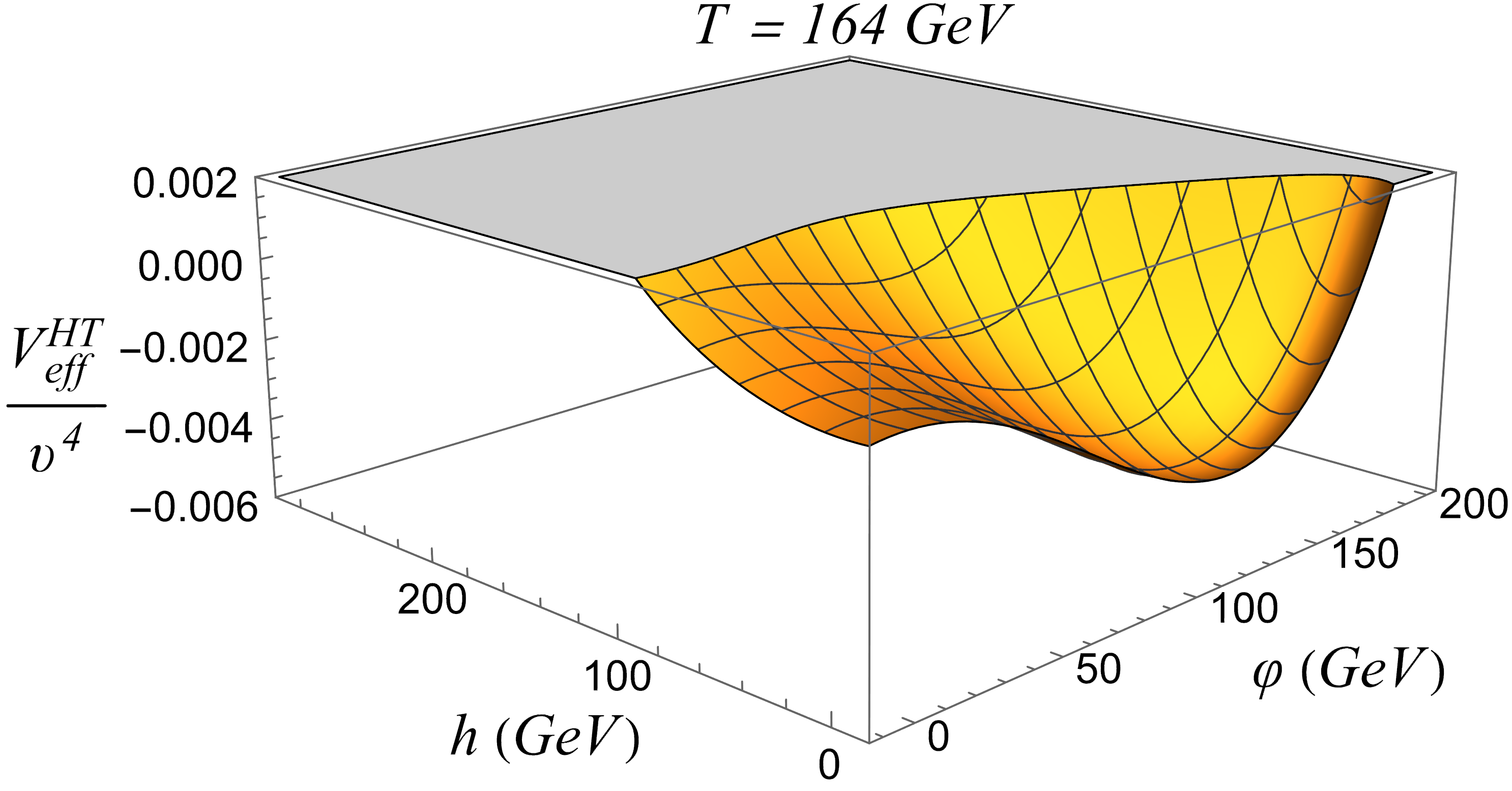}
\includegraphics[width=14pc]{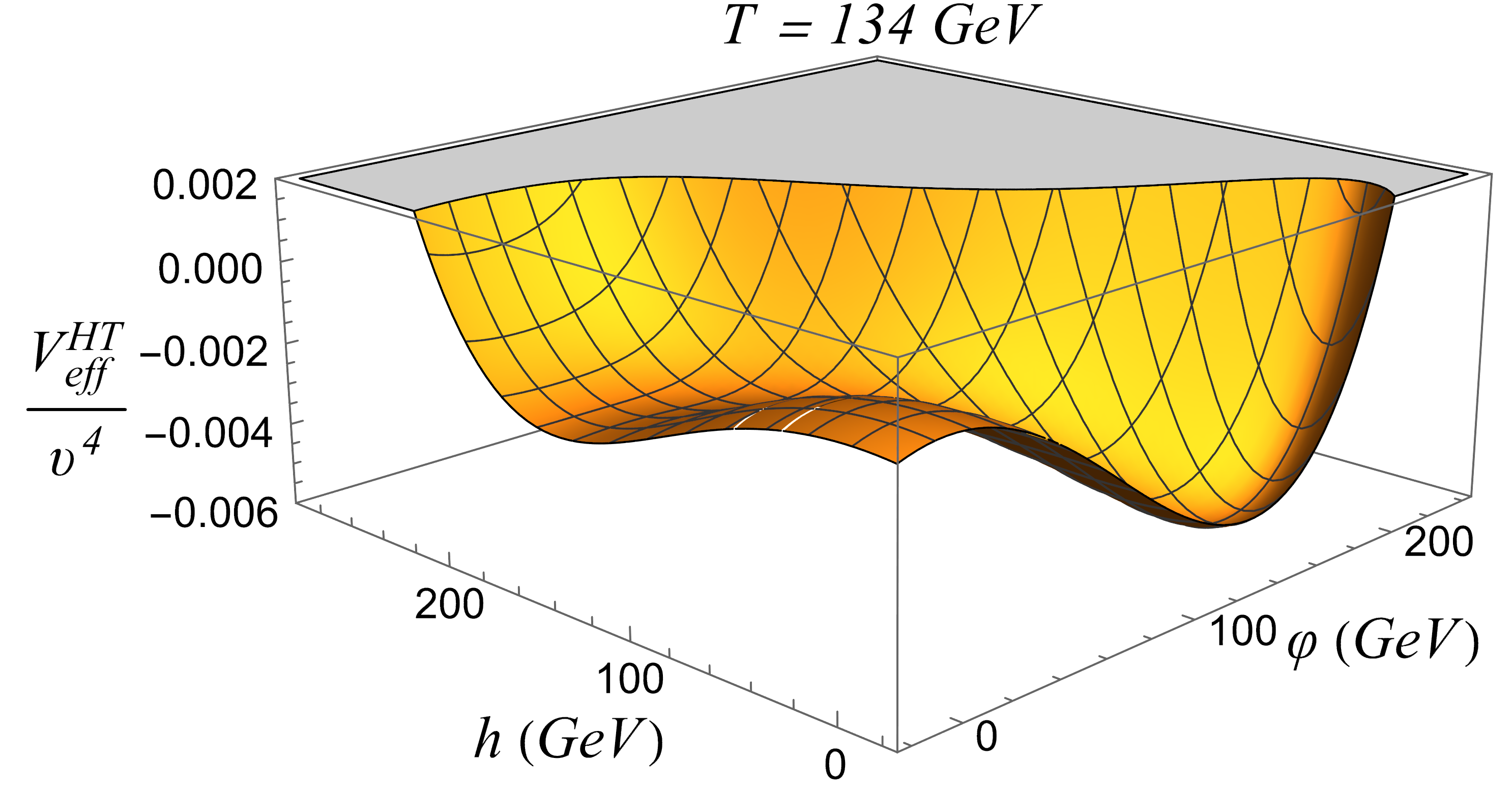}
\includegraphics[width=14pc]{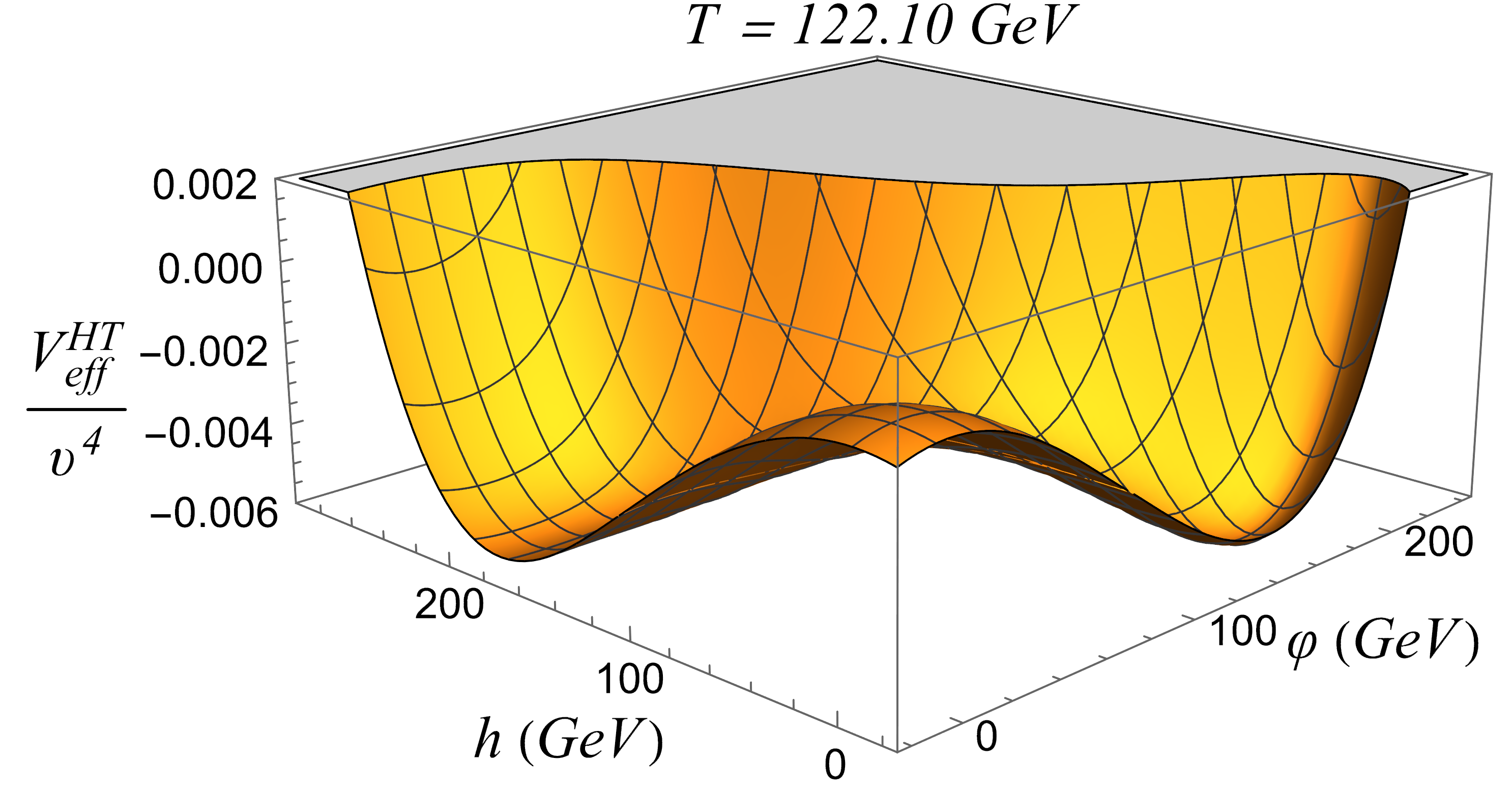}
\includegraphics[width=14pc]{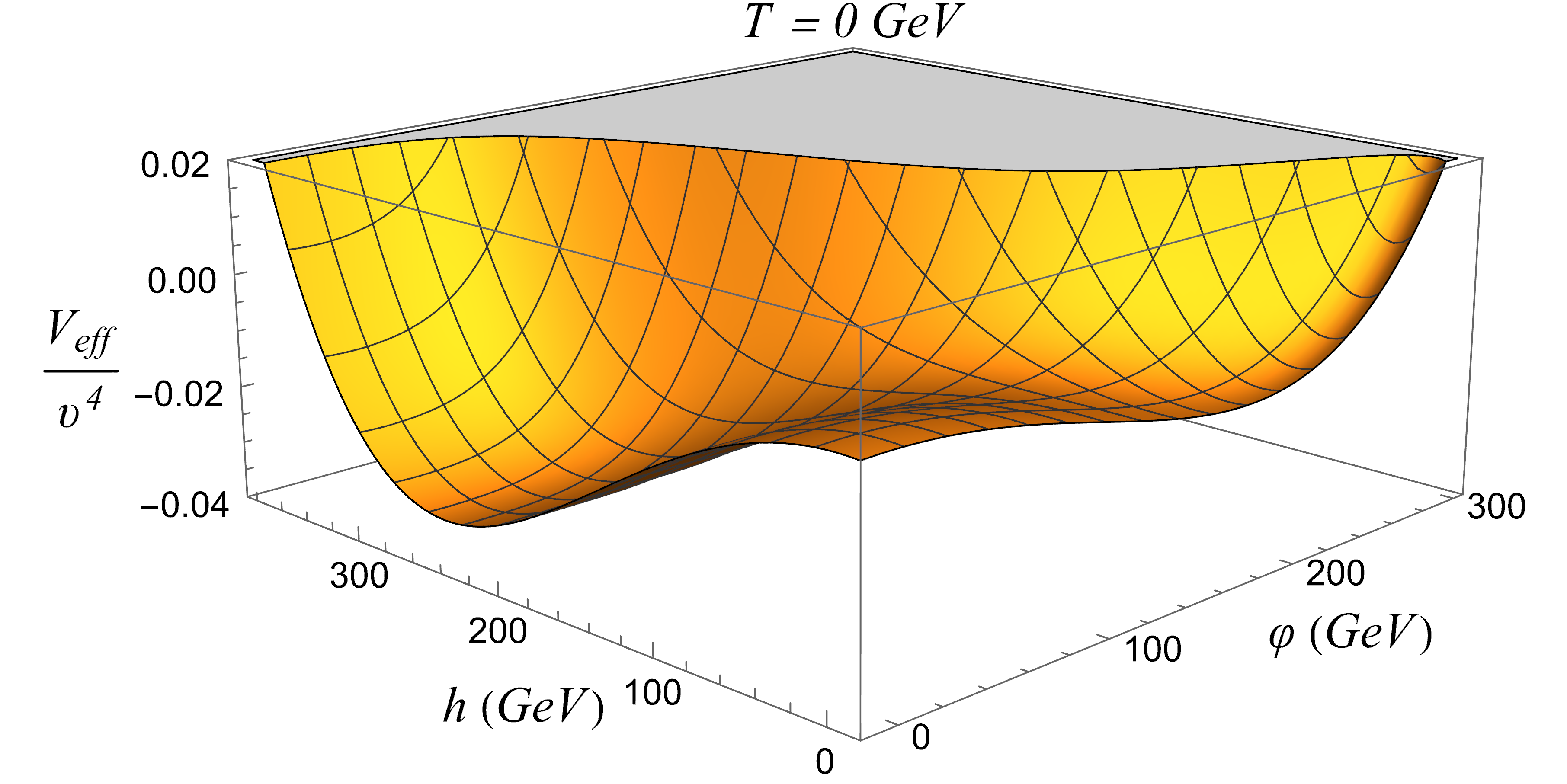}
\caption{The effective potential as the Universe cools down. In this example, the singlet's phase transition is second-order using a
point of the parameter space with \(m_S = 62.5\) GeV,
\(\lambda_{HS} = 0.15\), \(\lambda/M^2 \simeq 2 \times 10^{-5}\)
GeV\(^{-2}\), and \(a = 0.1\).} \label{T-potential_1}
\end{figure}

As it was mentioned earlier, the SM does not predict a strong first-order phase transition which is a necessary ingredient for successful EWBG. However, the corrections from the singlet extensions could assist the electroweak phase transition and explain the observed baryon asymmetry of the Universe. First of all, in the singlet extensions, different types of phase transitions can be realized, depending on the parameters of the real singlet extension. In particular, the sign of \(\mu_S^2\) mainly determines the different types of phase transitions. If \(\mu_S^2 > 0\), one notices that the electroweak phase transition is a two-step phase transition. Namely, the vacuum of the Universe initially resides at the origin \((h, \phi) = (0,0)\), and as the temperature decreases, a non-zero VEV appears in the \(\phi\) direction, which becomes the global minimum of the effective potential and a phase transition starts from the origin to the new global minimum \((0, \upsilon_s')\). When the Universe cools down enough, a local minimum in the \(h\) direction is formed, while a barrier separates this local minimum and the global minimum in the \(\phi\) direction. Below a critical temperature, the Higgs minimum becomes the global minimum, and a first-order phase transition occurs from \((0, \upsilon_s'')\) to \((\upsilon_c, 0)\), which are the minima at the critical temperature in the \(h\) and \(\phi\) direction, respectively. This two-step phase transition is illustrated in Figs. \ref{T-potential_2} and \ref{T-potential_1} for different singlet masses. One observes that the singlet's phase transition is a first-order phase transition for \(m_S = 500\) GeV in Fig. \ref{T-potential_2}, although it can also be a second order for lower singlet masses \(m_S > m_H/2\) and Higgs-singlet couplings \cite{Oikonomou:2024jms}. In contrast, the singlet's phase transition is generally second order in the parameter space with \(m_S = m_H/2\), as shown in Fig. \ref{T-potential_1}.

\begin{figure}[h!]
\centering
\includegraphics[width=15pc]{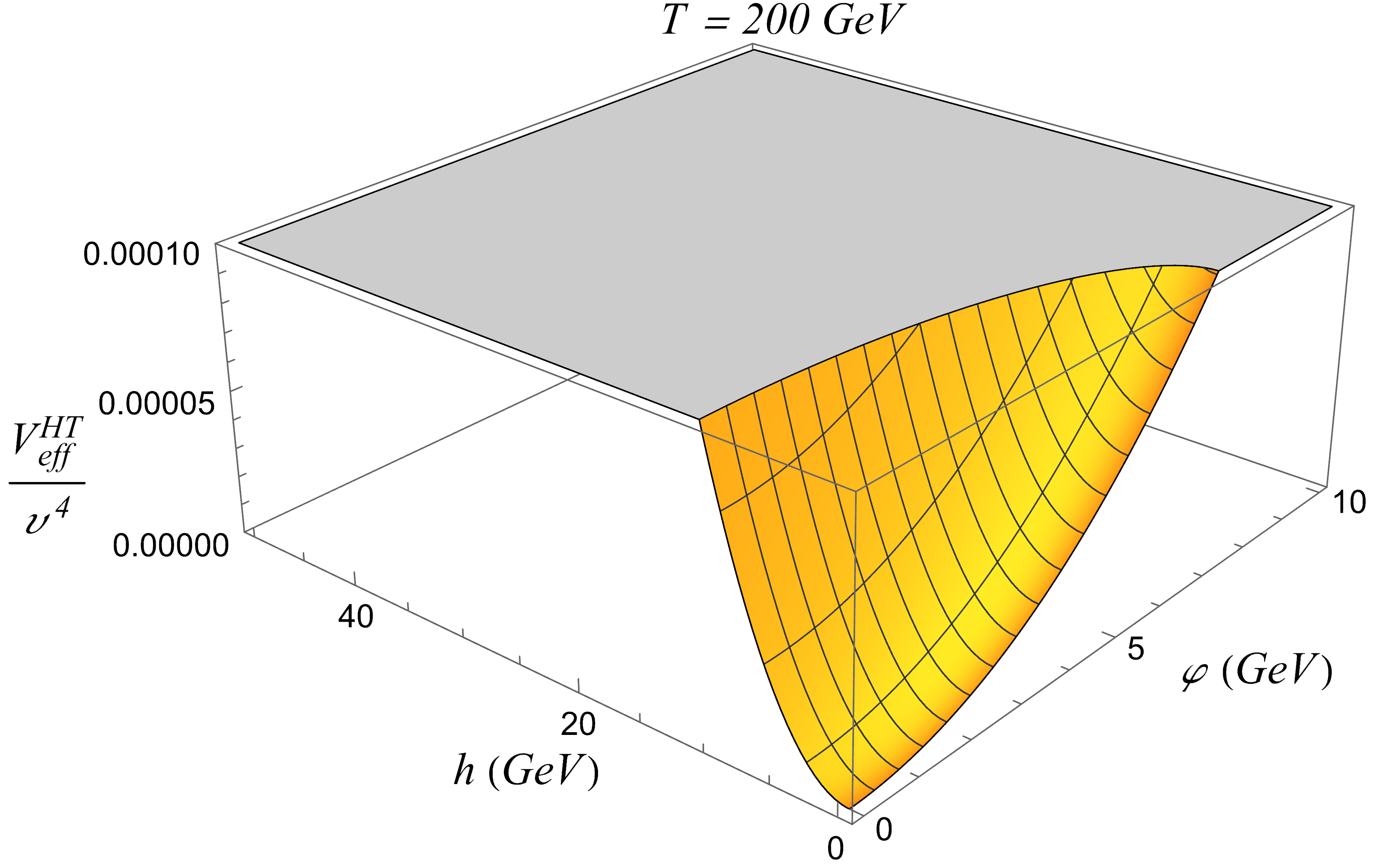}
\includegraphics[width=15pc]{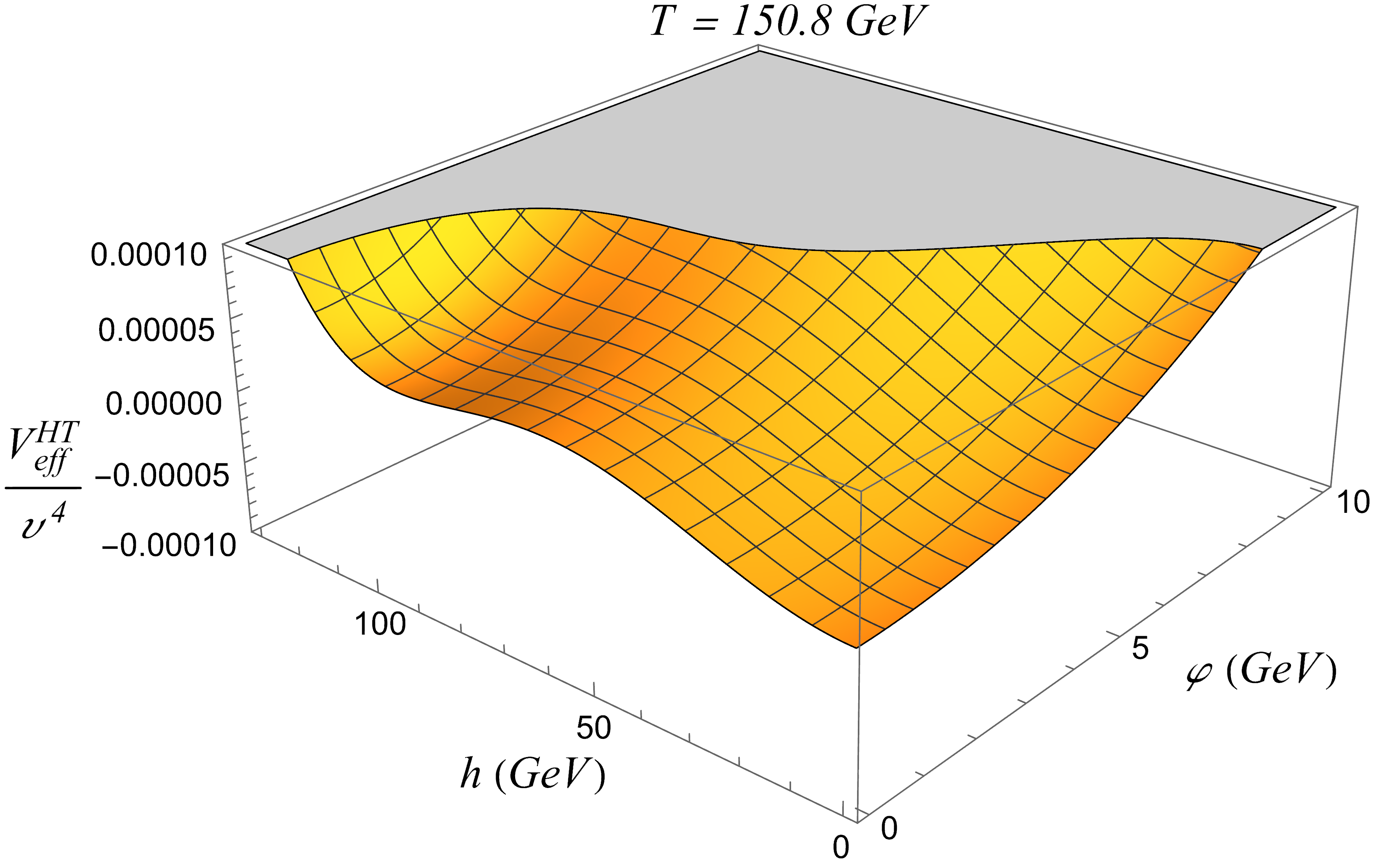}
\includegraphics[width=15pc]{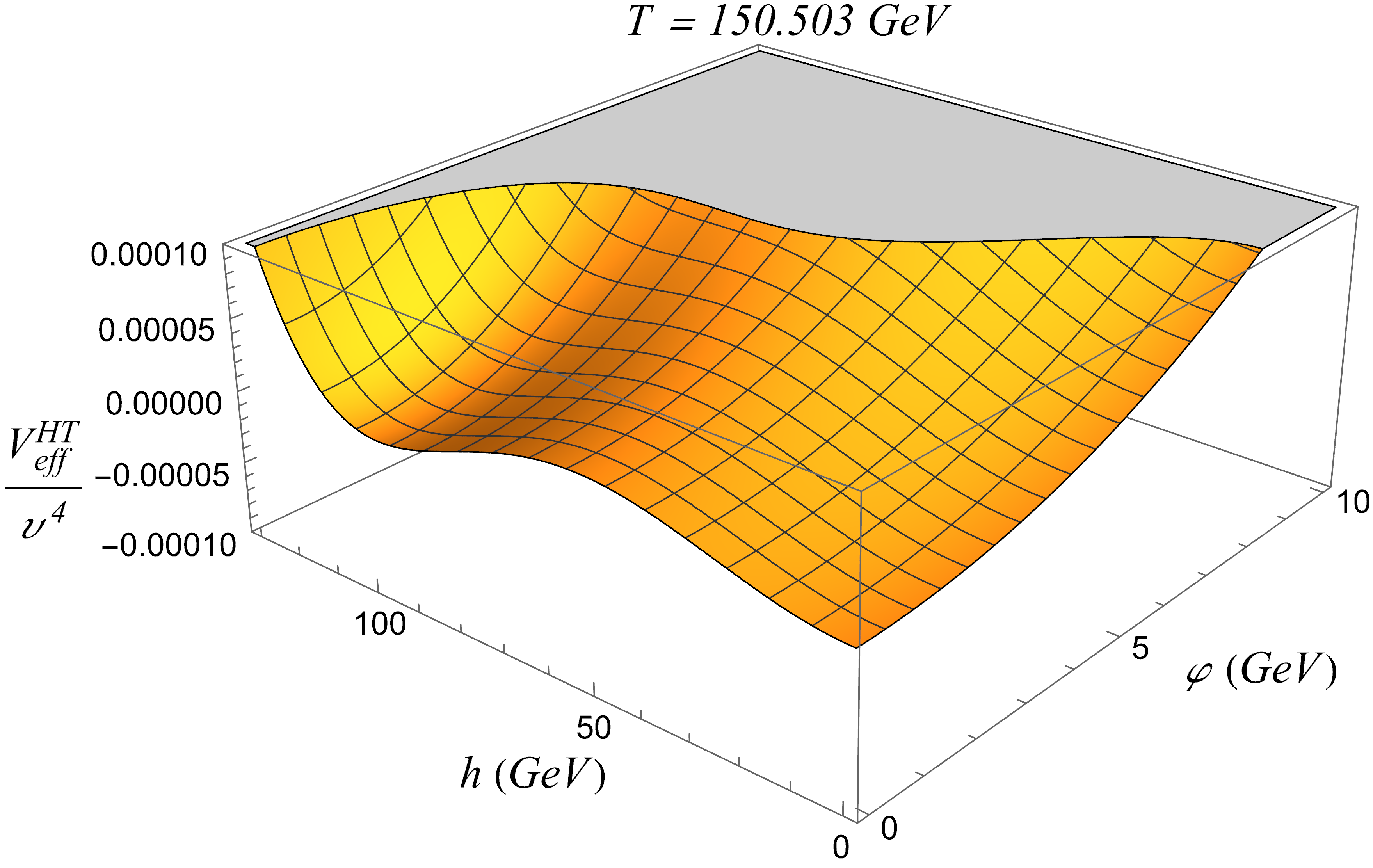}
\includegraphics[width=15pc]{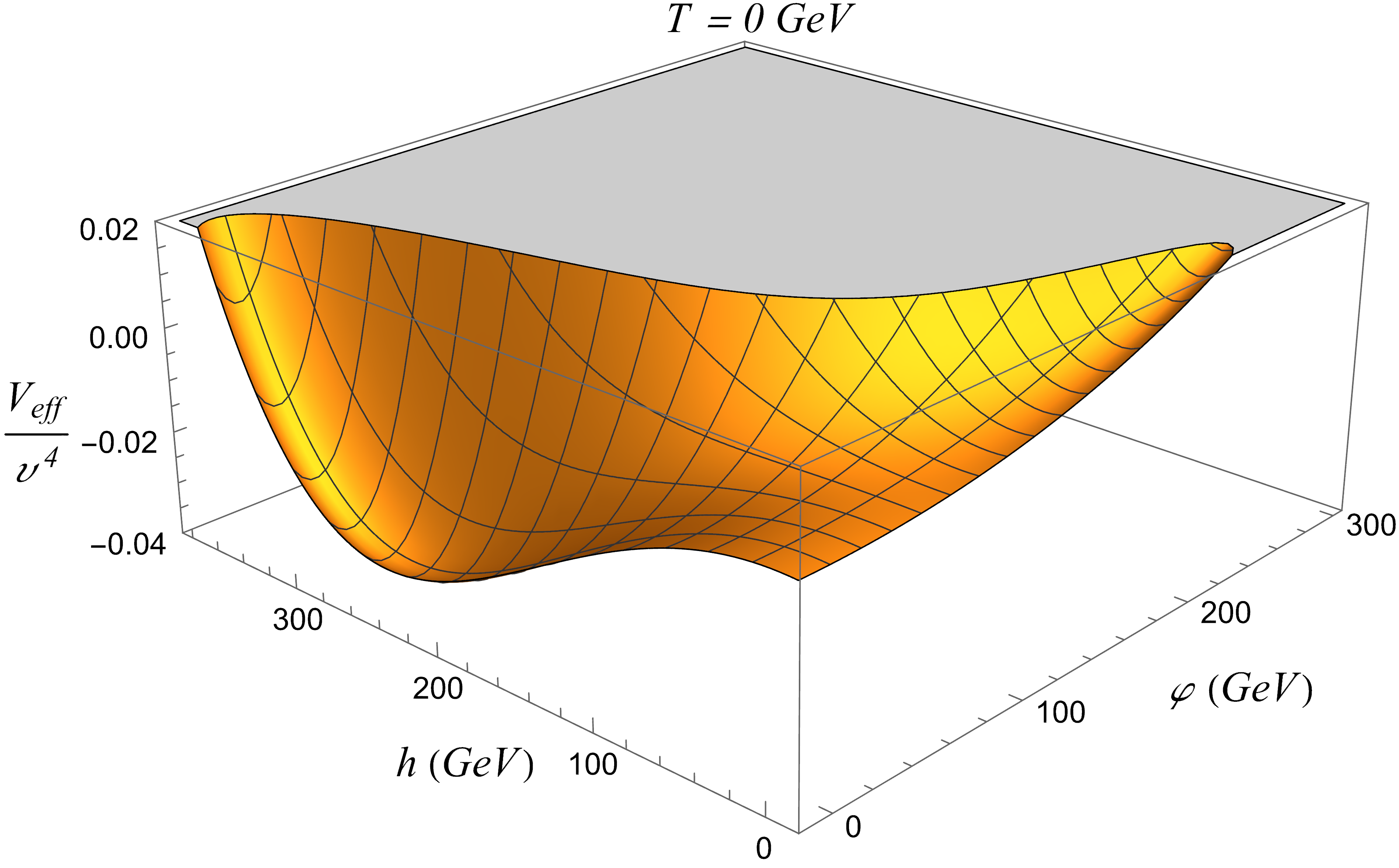}
\caption{The one-loop finite-temperature effective potential using a point of the parameter space with \(m_S = 400\) GeV, \(\lambda_{HS} = 2.6\), \(\lambda_S \simeq 0\), and \(\lambda = 0\). In this scenario, the critical temperature is around \(T_c = 150.503\) GeV.} \label{T-potential_mu_negative}
\end{figure}

It is essential to understand that in this region of the parameter space, the two-step electroweak phase transition occurs via tree-level effects\footnote{In fact, if \(\mu_S^2 > 0\), a one-step electroweak phase transition could also occur due to thermal effects. However, it was computed that this phase transition happens in a very limited region of the parameter space with \(\mu_S \lesssim 90\) GeV \cite{Oikonomou:2024jms}.} since the Higgs potential is extended through the singlet sector and the potential barrier, which appears between the Higgs and singlet minimum at the critical temperature, is also present at the tree level (Fig. \ref{treelevelpotential}). This barrier ensures the existence of the first-order electroweak phase transition, which may be strong depending on the parameters of the model. This scenario will be developed further in the next subsections as a viable EWBG occurs for large regions of the parameter space and the higher-order operator in the singlet extensions assists the strong electroweak phase transition in the majority of the parameter space. On the other hand, if \(\mu_S^2 < 0\), the electroweak phase transition is a one-step phase transition driven by loop effects. The singlet VEV remains at the origin \((h, \phi) = (0,0)\) as the temperature decreases and no tree-level effects assist the phase transition. The one-step electroweak phase transition evolves similarly to the SM case, but now the one-loop zero-temperature corrections coming from the singlet can induce a strong first-order phase transition as shown in Fig. \ref{T-potential_mu_negative}. In practice, the constraint for a successful EWBG (\ref{sphaleron_rate}) highly eliminates the parameter space with \(\mu_S^2 < 0\). In the singlet extensions with zero Wilson coefficient, a strong electroweak phase transition requires \(\lambda_{HS} \gtrsim 2\) and \(m_S \gtrsim 400\) GeV. The strong electroweak phase transition is also enhanced for quartic couplings \(\lambda_S\) close to zero. Then, the critical temperature varies from \(T_c \simeq 120 - 180\) GeV. However, the higher-order operator in the singlet extensions weakens significantly the electroweak phase transition, eliminating completely the viable parameter space. Therefore, the following discussion will focus solely on the case of \(\mu_S^2 \geq 0\) as the real singlet extension with a higher-order operator has no interest in the case of \(\mu_S^2\). More details about the one-step phase transition in the singlet extensions with \(\lambda = 0\) can be found in Ref. \cite{Curtin:2014jma}.

Before we analyze the parameter space of the real singlet extension with dimension-six operators, which describes a two-step electroweak phase transition, it is crucial to demonstrate with simple mathematical arguments the impact of the higher-order operator on the effective potential. Along the \(h\) direction, the non-zero Wilson coefficient modifies the singlet’s temperature-dependent self-energy, where the Wilson coefficient is compared with the Higgs-singlet coupling and the singlet's self-coupling, which then depends on the singlet mass, the Higgs-singlet coupling, and the parameter \(a\). This indicates that the influence of the higher-order operator is highly determined by the singlet mass and the Higgs-singlet coupling as it will be illustrated later. In contrast, along the \(\phi\) direction, the Wilson coefficient primarily affects the effective mass of the Higgs boson, the Goldstone bosons, and the singlet. As a result, the non-zero Wilson coefficient significantly contributes to the effective potential along the \(\phi\) direction, while large values of the Wilson coefficient affect the behavior of the effective potential in the \(h\) direction as well.

In the next subsections, the two-step strong electroweak phase transition is studied by dividing the parameter space of the extended SM into three regions: the low-mass region (\(m_S < m_H/2\)), and the high-mass region \((m_S \geq m_H/2\)).

\subsection{High-mass Region}

\begin{figure}[h!]
\centering
\includegraphics[width=30pc]{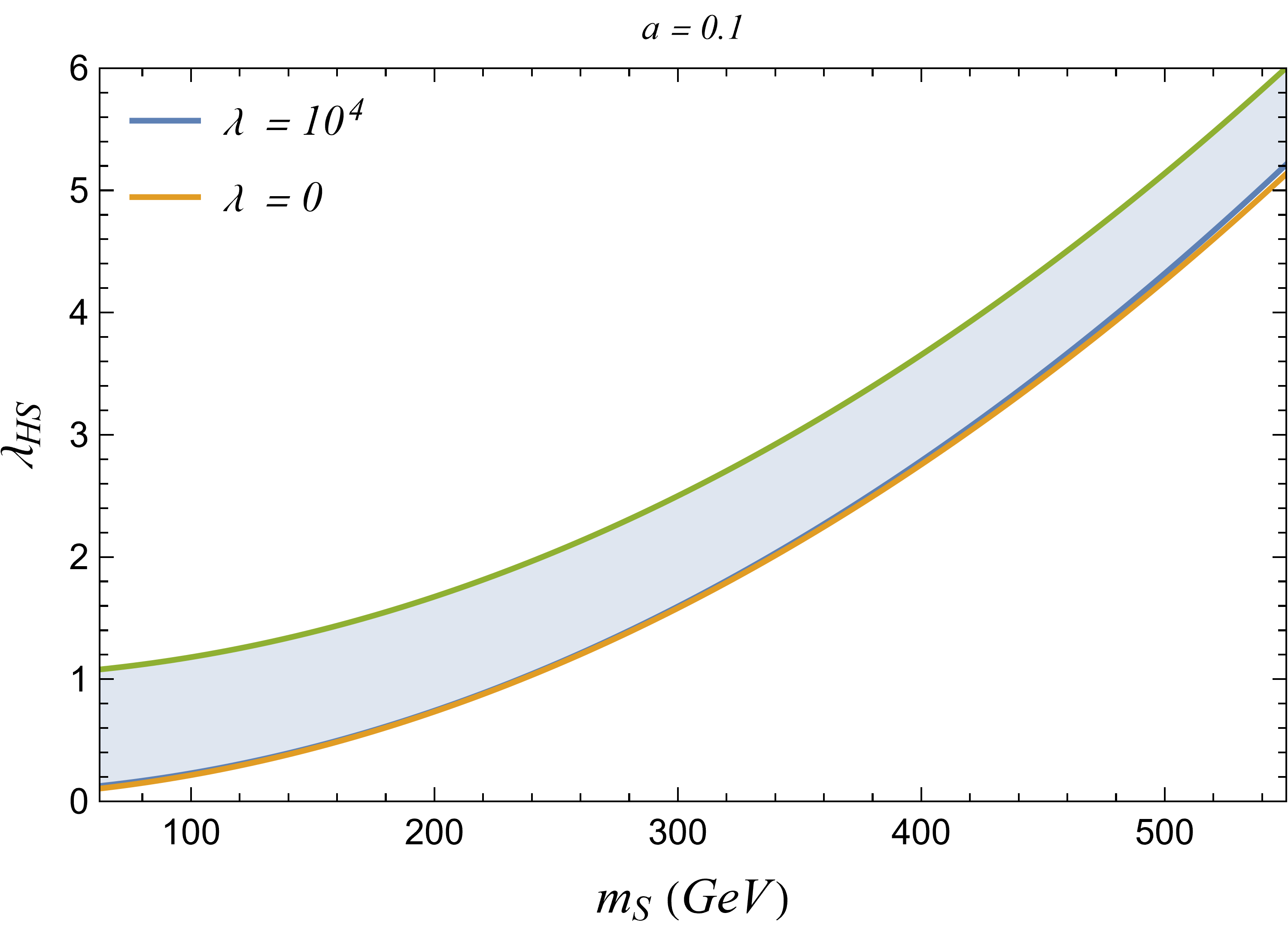}
\caption{The parameter space (blue region) of the singlet extension with a dimension-six operator (\(\lambda = 10^4\)) to realize a strong electroweak phase transition (\(\upsilon_c/T_c > 1\)) using \(a = 0.1\). The orange line shows the lower bound of the parameter space of the singlet extension without a dimension-six operator (\(\lambda = 0\)). The critical temperature in this scenario ranges from \(T_c \simeq 30 - 140\) GeV.}\label{ParameterSpace2}
\end{figure}

Firstly, the parameter space for \(m_S > m_H/2\) and \(\lambda = 0\) allows us to realize a two-step electroweak phase transition, choosing the usual parametrization with \(a = 0.1\). However, the condition (\ref{sphaleron_rate}) for a strong first-order phase transition eliminates this parameter space as mentioned in the previous section. As a result, the parameter space for a strong two-step electroweak phase transition with \(\lambda = 0\) is illustrated in Fig. \ref{ParameterSpace2} bounded by the orange line. In general, these results are consistent with previous studies \cite{Curtin:2014jma, Chiang:2018gsn, Beniwal:2017eik,Jain:2017sqm}, although minor differences may arise between our results and those in the literature due to the theoretical uncertainties in the perturbative analysis, as shown in Refs. \cite{Chiang:2018gsn, Athron:2022jyi, Croon:2020cgk, Gould:2021oba}.

\begin{figure}[h!]
\centering
\includegraphics[width=30pc]{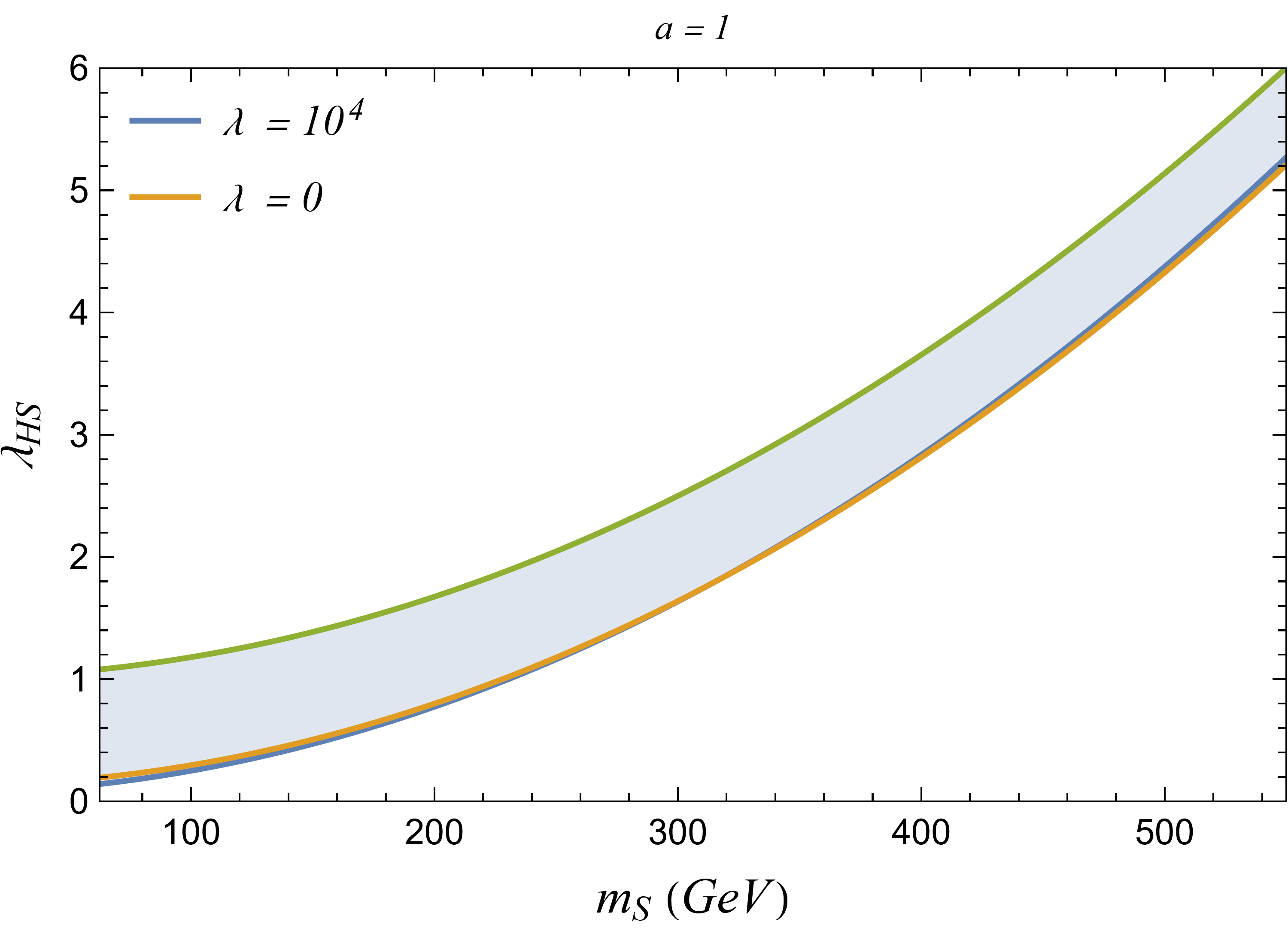}
\caption{The parameter space (blue region) of the singlet extension with a dimension-six operator (\(\lambda = 10^4\)) to realize a strong electroweak phase transition (\(\upsilon_c/T_c > 1\)) using \(a=1\). The orange line shows the lower bound of the parameter space of the singlet extension without a dimension-six operator (\(\lambda = 0\)).}\label{ParameterSpace3}
\end{figure}

In the singlet extension with a dimension-six operator, the parameter space does not change considerably for a Wilson coefficient \(\lambda < 10^{2}\). If the Wilson coefficient is much larger, the lower bound on the Higgs-singlet coupling increases, whereas the ratio \(\upsilon_c/T_c\) decreases, especially for low and very high singlet masses. In Fig. \ref{ParameterSpace2}, the parameter space with \(\lambda = 0\) and \(\lambda = 10^4\) are compared to show that a non-zero Wilson coefficient reduces the parameter space for \(a = 0.1\). In contrast, if we set \(a = 1\), the higher-order operator assists the strong electroweak phase transition for low singlet masses, which means that a lower Higgs-singlet coupling induces a strong phase transition for a non-zero Wilson coefficient. This is clearly demonstrated in Fig. \ref{ParameterSpace3}. This behavior is more evident for different values of the parameter \(a\) and lower singlet masses as shown in the next paragraphs.

\begin{figure}[h]
\centering
\includegraphics[width=20.5pc]{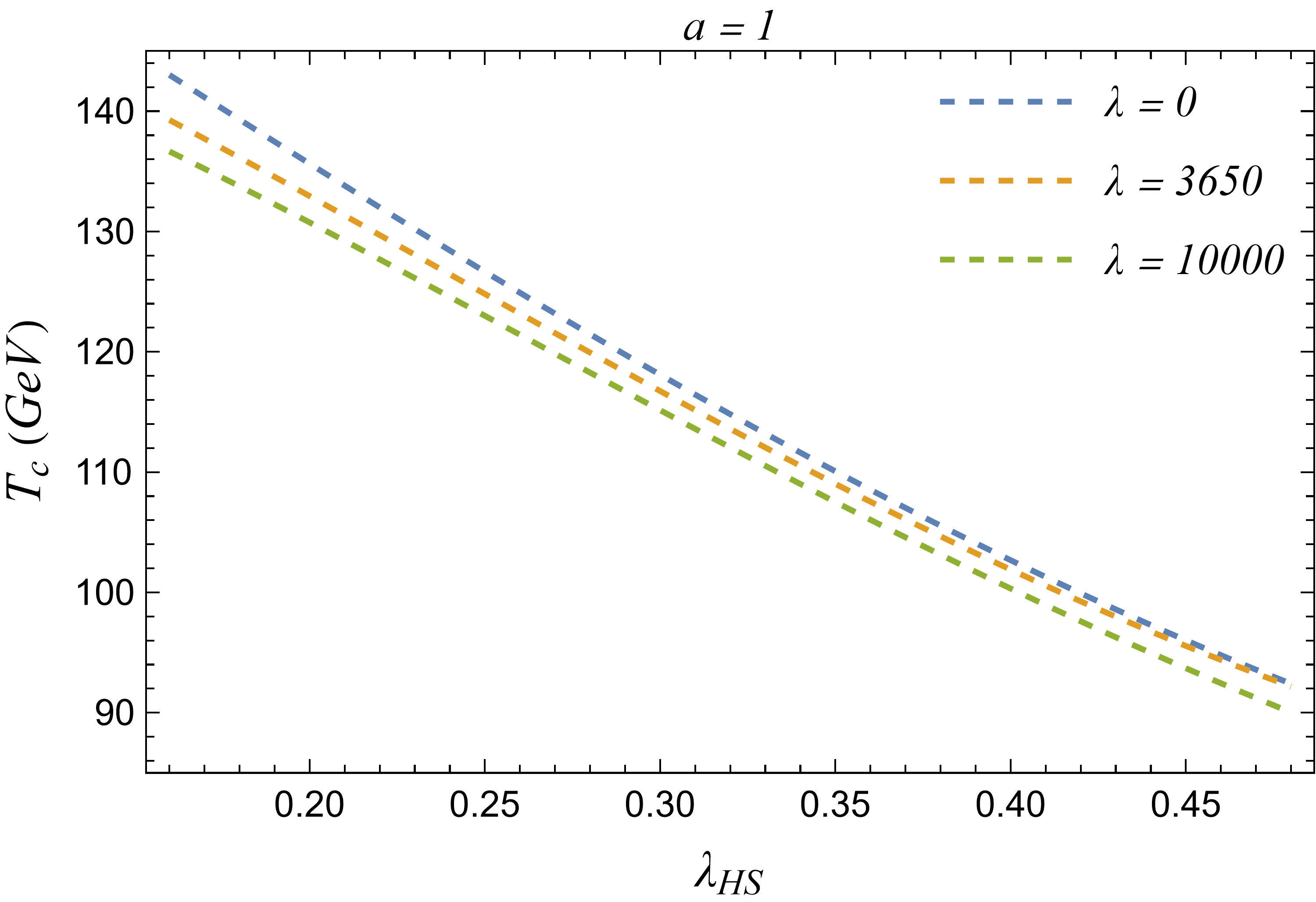}
\includegraphics[width=20.5pc]{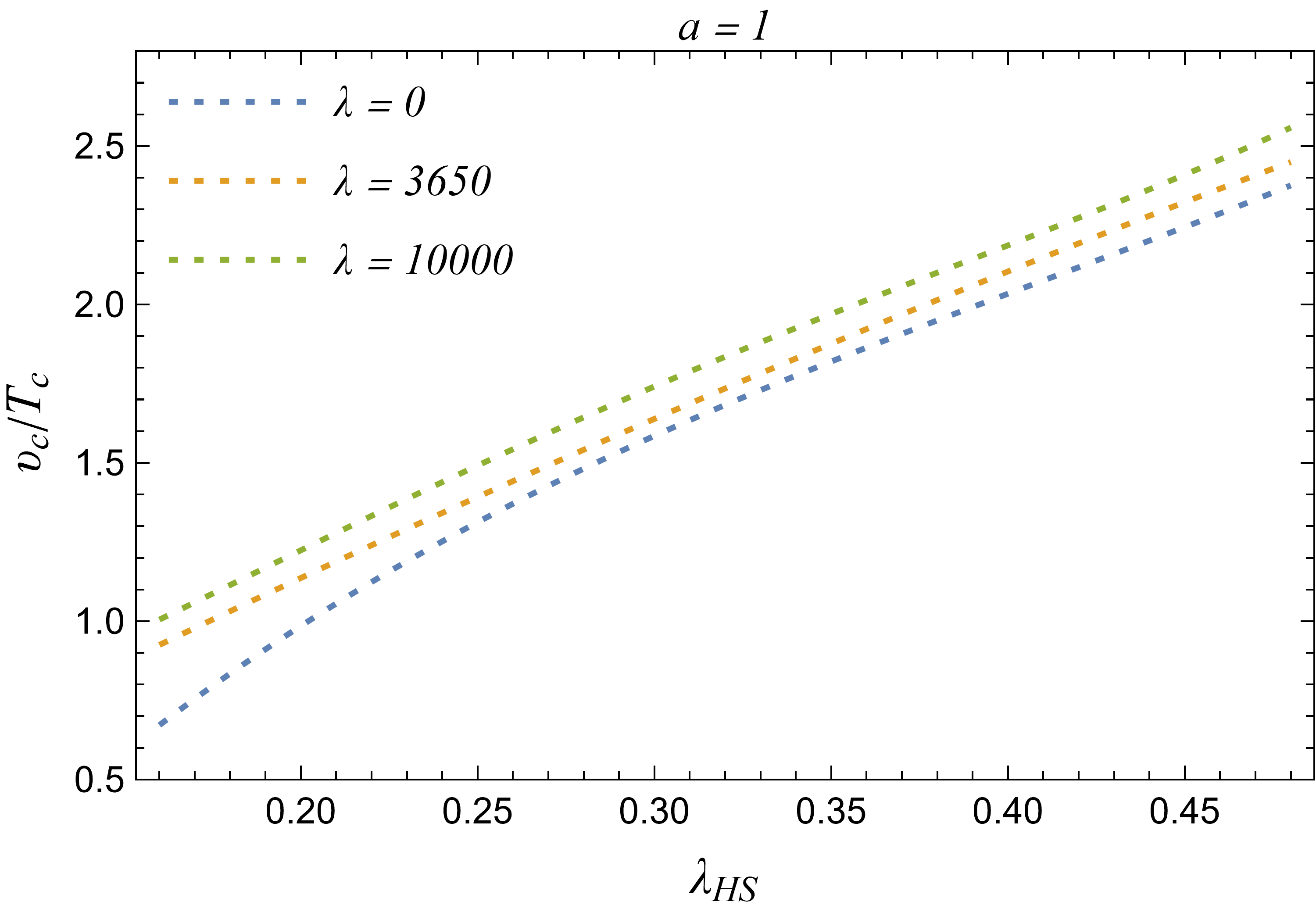}
\caption{The electroweak phase transition with $m_S = 62.5 $ GeV and \(a = 1\): The critical temperature (left) and the ratio in sphaleron rate (right) as a function of the Higgs-singlet coupling for various values of the Wilson coefficient.}\label{Tc_sr_ms_625_a_1}
\end{figure}

In the Higgs resonance region \(\left(m_S = m_H/2\right)\), the parameter space for a strong electroweak phase transition is expanded by taking \(a \gtrsim 0.4\) owing to the presence of the higher-order operator \cite{Oikonomou:2024jms}. This is obvious in the case of \(a = 1\) in Fig. \ref{Tc_sr_ms_625_a_1}, where the constraint \(\upsilon_c/T_c > 1\) can be satisfied by much lower \(\lambda_{HS}\) compared to the case with zero Wilson coefficient. Namely, the higher-order operator could assist a strong electroweak phase transition in regions of the parameter space that were previously excluded in the literature. As a consequence, these findings suggest that dimension-six operators in the real singlet extension provide a viable framework for EWBG, expanding the viable parameter space in the Higgs resonance region for testing EWBG through experimental signatures, especially at gravitational wave experiments and dark matter searches \cite{GAMBIT:2017gge, Athron:2018ipf, Feng:2014vea}.

The dependence of the critical temperature on the Higgs-singlet coupling in the Higgs resonance region with \(a = 1\) is illustrated in Fig. \ref{Tc_sr_ms_625_a_1} (left). It is evident that the impact of the higher order operator is noticeable for Wilson coefficients \(\lambda > 10^3\) since the critical temperature with a lower Wilson coefficient tends toward the critical temperature with zero Wilson coefficient for larger values of the Higgs-singlet coupling. This behavior showcases that the impact of the higher-order operator is stronger for low Higgs-singlet couplings. Similarly, this trend is evident in Fig. \ref{Tc_sr_ms_625_a_1} (right), which shows the ratio \(\upsilon_c/T_c\) in the condition (\ref{sphaleron_rate}) for a successful EWBG.

\subsection{Low-mass Region}

\begin{figure}
\centering
\includegraphics[width=21.25pc]{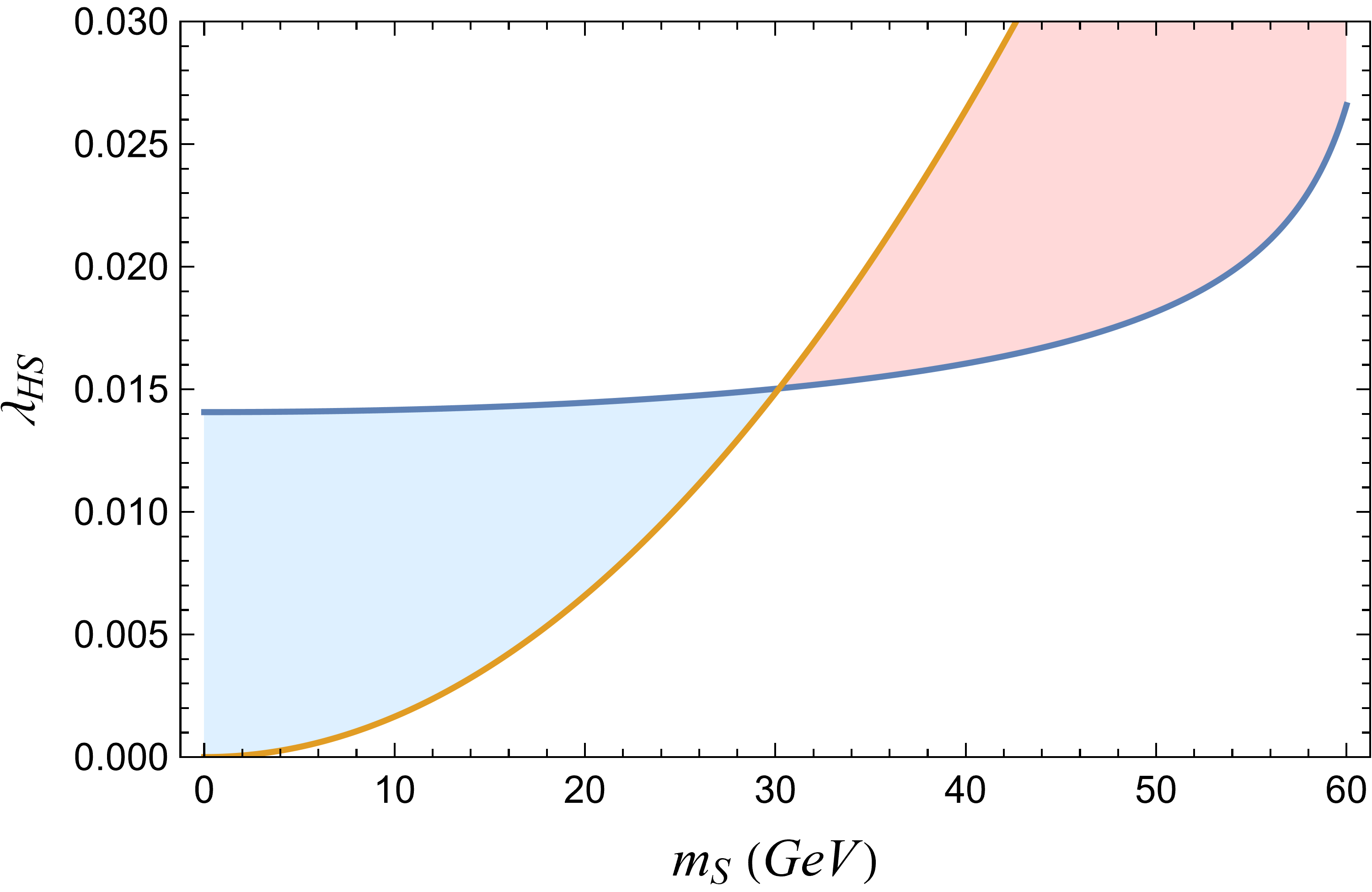}
\includegraphics[width=21pc]{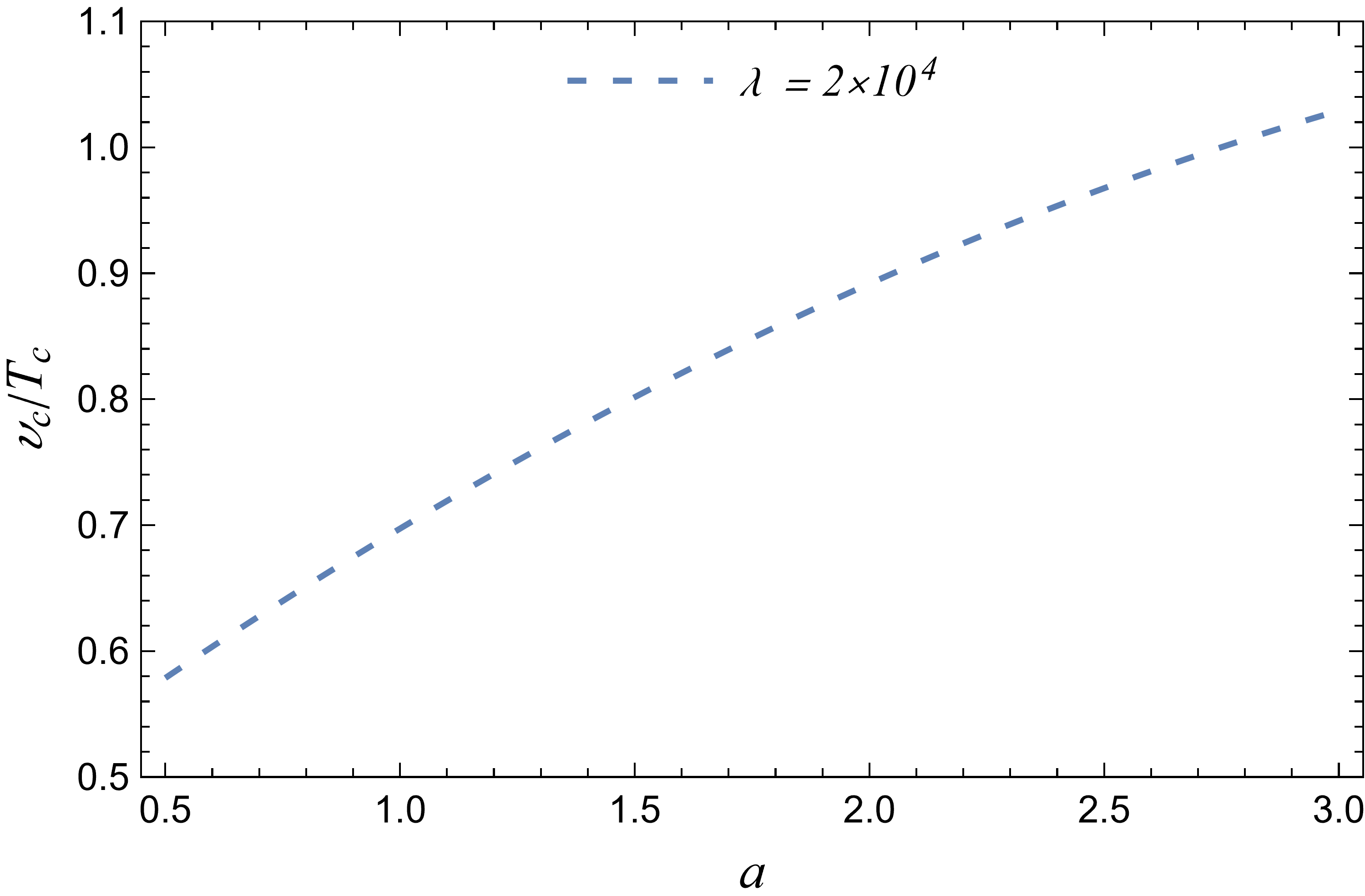}
\caption{\textbf{Left}: The parameter space (blue region) of the singlet extension with \(\lambda \geq 10^3\), \(a \geq 0.75\), and \(m_S \leq 30\) GeV, where a strong electroweak phase transition (\(\upsilon_c/T_c > 0.6\)) occurs. This two-dimensional parameter space \((m_S, \lambda_{HS})\) corresponds to the following points \((\lambda, a)\): \((0.75, 2 \times 10^4)\), \((1.00 , 15\times 10^3)\), \((1.25, 15 \times 10^3)\), and \((1.75, 10^3)\). The blue line bounds the parameter space due to the constraint by the invisible Higgs decay width (\ref{special_condition_coupling}) and the orange line represents \(\mu_S^2 > 0\). \textbf{Right}: The ratio in the sphaleron rate \(\left(\upsilon_c/T_c\right)\) as a function of \(a\) for \(m_S \leq 1\) GeV, \(\lambda_{HS} \leq 16 \times 10^{-4}\), and \(\lambda = 2 \times 10^4\). }\label{Parameter_Space_ms_lower_than_30}
\end{figure}

In the low-mass region, the parameter space is highly restricted by the constraint from the invisible Higgs decay width\footnote{The branching ratio of the Higgs boson to invisible particles is set to \(BR_{\text{inv}} < 0.19\).}. More specifically, if \(a \gtrsim 0.05 \), in the singlet extensions without a higher-order operator, the entire lower mass region (\(m_S \lesssim 30\) GeV) in Fig. \ref{BR_0.19} is ruled out by imposing the condition for a successful EWBG (\ref{sphaleron_rate}). In contrast, large values of the Wilson coefficient with \(a \gtrsim 0.05 \) can assist the strong phase transition in the lower mass region \cite{Oikonomou:2024jms}. This is clearly demonstrated in Fig. \ref{Parameter_Space_ms_lower_than_30} (left), where a strong phase transition (\(\upsilon_c/T_c > 0.6\)) occurs for every coupling \(\lambda_{HS}\) and singlet mass constrained by (\ref{condition_coupling}), taking various points \((\lambda, a)\) in the four-dimensional parameter space. Smaller values of the Wilson coefficient require larger values of \(a\) to realize a strong electroweak phase transition. Fig. \ref{Parameter_Space_ms_lower_than_30} (right) showcases how the strength of the electroweak phase transition increases, while the parameter \(a\) increases for a constant Wilson coefficient and the parameter space \(\left(m_S, \lambda_{HS} \right)\) remains the same as in Fig. \ref{Parameter_Space_ms_lower_than_30} (right), which has the maximum size allowed by (\ref{condition_coupling}). The strength of the electroweak phase transition increases further for higher singlet masses and Higgs-singlet couplings. Thus, a strong electroweak phase transition can be generated for low singlet masses \(m_S < m_H/2\) by including the dimension-six operator with \(\lambda/M^2 > 5 \times 10^{-6}\) GeV\(^{-2}\) and \(a > 0.05\) \cite{Oikonomou:2024jms}. In contrast, if \(a \lesssim 0.05\), this behavior shifts, since the inclusion of the higher-order operator does not assist the electroweak phase transition. Thus, the results for \(a > 0.05\) provide new pathways for achieving a strong electroweak phase transition, expanding the viable parameter space of low singlet masses for experimental exploration at future collider and gravitational wave experiments.

In the higher mass region of the parameter space in Fig. \ref{BR_0.19}, the singlet mass is close to half of the Higgs mass as it is constrained by the condition in  (\ref{condition_coupling}). Consequently, the behavior of the effective potential in the higher mass region does not differ significantly from the one in the Higgs resonance region, while the results on the critical temperature and the Higgs VEV at the critical temperature are similar to the results in the previous subsection.

\section{Conclusions and Discussion}

In this article, we have investigated the EWBG in the real singlet extension of the SM, introducing a real singlet scalar field, which couples through the Higgs portal, imposing a \(\mathbb{Z}_2\) symmetry. We also assumed that the singlet scalar field couples to the Higgs boson via a dimension-six operator, coming from an effective field theory active at an energy scale \(M= 15 - 100\) TeV. Then, the parameter space of this extension is defined by the four parameters \(\left(\mu_S^2, \lambda_{HS}, \lambda_S, \lambda \right)\) or \(\left(m_S, \lambda_{HS}, a, \lambda \right)\) and is primarily restricted by the condition for a strong electroweak phase transition, the vacuum structure, and the validity of the one-loop perturbative analysis. The parameter space is reduced further, considering the constraints by the experimental signatures of EWBG in the real singlet extension, such as the invisible Higgs decays. Particle colliders can also probe the modifications to the triple Higgs coupling and the \(Zh\) cross section due to the singlet's one-loop corrections \cite{Curtin:2014jma, Beniwal:2017eik}. We then discussed the stochastic gravitational wave background generated by a first-order electroweak phase transition \cite{Kosowsky:1991ua, Kamionkowski:1993fg, Huber:2008hg, Jinno:2016vai, Caprini:2007xq, Kosowsky:1992vn, Kosowsky:1992rz,Hindmarsh:2013xza, Hindmarsh:2015qta, Giblin:2013kea, Giblin:2014qia,Kahniashvili:2008pe, Kahniashvili:2009mf, Caprini:2009yp, Binetruy:2012ze}, which can be detected by future gravitational wave observations and is one of the most promising experimental signatures of EWBG in the singlet-extended SM. 

Initially, the parameter space is separated into two regions, depending on the sign of \(\mu_S^2 \) since a strong electroweak phase transition occurs as a two-step phase transition for \(\mu_S^2 > 0\) via tree-level effects and as a one-step phase transition for \(\mu_S^2 < 0\) due to the one-loop zero-temperature corrections, coming from the singlet sector. During a strong one-step phase transition, the Higgs boson acquires a non-zero VEV, while the singlet VEV remains zero. This phase transition is mainly observed in the parameter space with \(m_S > 400\) GeV, \(\lambda_{HS} > 2\), \(\lambda_S \simeq 0\), and \(\lambda = 0\), whereas a non-zero Wilson coefficient weakens significantly the electroweak phase transition. On the other hand, the two-step electroweak phase transition consists of an initial singlet phase transition at high temperatures and a subsequent strong first-order phase transition from the singlet to the Higgs vacuum. A two-step phase transition is by far more interesting in cosmology and particle physics and it was studied in more detail in order to comply with all the theoretical and experimental constraints for \(m_S = 0 - 550\) GeV.

If \(\mu_S^2 > 0\), the parameter space is then divided into the high-mass region \(\left(m_S \geq m_H/2\right)\) and the low-mass region \(\left(m_S < m_H/2\right)\). Firstly, the parameter space in the high-mass region for \(a \gtrsim 0.4\) is expanded for large values of the Wilson coefficient, although this trend is reversed for \(a < 0.4\). However, the presence of the dimension-six operator does not affect significantly the parameter space of the usual singlet extension for very high singlet masses. On the other hand, the presence of the dimension-six operator with \(\lambda/M^2 \gtrsim 5 \times 10^{-6}\) GeV\(^{-2}\) and \(a \gtrsim 0.05\) can assist a strong electroweak phase transition for low singlet masses \(m_S < m_H/2\). The low-mass region with \(\lambda = 0\) is severely eliminated by the constraints from the invisible Higgs decay searches and the condition for a successful EWBG. Consequently, a dimension-six operator can induce strong electroweak phase transition for low Higgs-singlet couplings, which were not allowed in the usual singlet extensions of the SM, and low singlet masses, which were excluded by the invisible Higgs decay searches.

Except for the departure from thermal equilibrium, EWBG requires a viable source of \(CP\)-violation. In the real singlet extension, this source is introduced by a dimension-six operator that couples the singlet to the top-quark mass and extends the top-quark Yukawa interaction term in the SM Lagrangian. This higher-order operator does not change considerably the finite-temperature effective potential as it is suppressed by a new physics scale beyond 1 TeV \cite{Cline:2012hg}. Subsequently, the top-quark mass acquires a complex phase, resulting in the \(CP\)-violation required to explain the baryon asymmetry of the Universe. This scenario is also consistent with considering the singlet as a dark matter candidate, with a mass close to half that of the Higgs boson, although it is constrained by the direct dark matter detection experiments \cite{GAMBIT:2017gge, Athron:2018ipf, Feng:2014vea}.

Overall, this work supports the feasibility of a strong electroweak phase transition in the real singlet extension with dimension-six operators, providing a framework for testing EWBG and expanding the theoretical and experimental landscape for future collider and gravitational wave searches.

\section*{APPENDIX A: Gravitational Waves from Cosmological Phase Transitions}

We introduce the details of the gravitational wave spectrum induced by a first-order electroweak phase transition. First-order phase transitions in the early Universe are violent processes that generate a stochastic gravitational wave background. In general, the main sources of gravitational waves from a first-order phase transition are the bubble collisions \cite{Kosowsky:1991ua, Kamionkowski:1993fg, Huber:2008hg, Jinno:2016vai, Caprini:2007xq, Kosowsky:1992vn, Kosowsky:1992rz}, the sound waves in the plasma after the phase transition \cite{Hindmarsh:2013xza, Hindmarsh:2015qta, Giblin:2013kea, Giblin:2014qia} and the magneto-hydrodynamical turbulence (MHD) in the plasma after bubble collisions \cite{Kahniashvili:2008pe, Kahniashvili:2009mf, Caprini:2009yp, Binetruy:2012ze}. In particular, a first-order phase transition involves the release of latent heat due to the bubble nucleation. While the bubble walls expand in the plasma of particles at high temperatures, they collide, and gravitational waves are emitted through the breaking of spherical symmetry in the bubble walls. Additionally, the expanding bubble walls accelerate the waves of plasma, which propagate along with the bubble wall, resulting in gravitational waves. Lastly, turbulence can be induced by the shocks in the fluid, leading to gravitational waves. 

A stochastic gravitational wave background is described by the spectral energy density,
\begin{equation}
    \Omega_{\text{GW}} (f) = \frac{1}{\rho_{\text{crit}}} \frac{d \rho_{\text{GW}}}{d \ln f}.
\end{equation}
In general, it is assumed as an approximation that the three main contributions to the gravitational wave spectrum can be added to compute the total energy density spectrum \cite{Caprini:2015zlo},
\begin{equation}
    h^2\Omega_{\text{GW}}  = h^2\Omega_{\text{col}}  +  h^2\Omega_{\text{sw}} + h^2\Omega_{\text{turb}} .
\end{equation}
If non-runaway bubbles are formed, during the first-order phase transition, the contribution of the bubble collisions can be neglected and the gravitational wave spectrum is mostly determined by 
\begin{equation}
    h^2\Omega_{\text{GW}}  \simeq h^2 \Omega_{\text{sw}}  + h^2\Omega_{\text{turb}} .
\end{equation}
The gravitational wave spectrum associated with a first-order phase transition is mainly computed using three parameters: the ratio of released latent heat from the transition to the energy density of the plasma background \(\alpha\), the inverse time of the phase transition \(\beta\), and the velocity of the bubble wall \(v_w\). The first important parameter is defined as 
\begin{equation}
    \alpha =\frac{1}{\rho_{\text{rad}}} \left. \left[ V_{\text{F}} - V_{\text{EW}} + T \left( \frac{d V_{\text{EW}}}{dT} - \frac{d V_{\text{F}}}{dT} \right)\right] \right|_{T = T_*},
\end{equation}
where \(T_*\) is the critical temperature when the gravitational waves are produced and \(V_{\text{EW}}\) and \(V_{\text{F}}\) are the potential energies in the true vacuum state \(\left(h, \phi\right) = \left( \upsilon, 0\right)\) and in the false vacuum state \(\left(h, \phi\right) = \left(0, 0\right)\), respectively \cite{Caprini:2015zlo}. Then, the inverse time duration of the phase transition reads
\begin{equation}
      \beta= H(T_*) T_* \left. \frac{d}{dT} \left(\frac{S_3}{T} \right) \right|_{T = T_*}.
\end{equation}
These key parameters depend on the three-dimensional action and the critical temperature. In the real singlet extension, the Euclidean action of a critical bubble is cast into the form,
\begin{equation}
    S_3 = 4 \pi \int r^2 dr\, \left[\frac{1}{2} \left(\frac{d h_b}{d r}\right)^2 + \frac{1}{2} \left(\frac{d \phi_b}{d r}\right)^2 + V_{\text{eff}}(h_b, \phi_b, T)  \right],
\end{equation}
where \((h_b, \phi_b)\) is the bubble profile of the critical bubble given by the bounce solution of the following equations of motion,
\begin{equation}
    \frac{d^2 h_b}{d r^2} + \frac{2}{r} \frac{d h_b}{d r} = \frac{\partial V_{\text{eff}}}{\partial h_b}, \quad \frac{d^2 \phi_b}{d r^2} + \frac{2}{r} \frac{d \phi_b}{d r} = \frac{\partial V_{\text{eff}}}{\partial \phi_b}.
\end{equation}
with the corresponding boundary conditions. These equations of motion can be solved following the method in Ref. \cite{Beniwal:2017eik}. Then, the critical temperature can be defined as the temperature at which the probability for a single bubble to be nucleated within one horizon volume approaches one \cite{Grojean:2006bp}. The dynamics of first-order phase transitions are presented further in Ref. \cite{Athron:2023xlk}. Therefore, it is shown in Ref. \cite{Huber:2008hg} that the peak frequency of the gravitational wave signals from bubble collisions is written as
\begin{equation}
    f_{\text{col}} = 16.5 \times 10^{-6} \frac{0.62}{v_w^2-0.1 v_w+1.8} \frac{\beta}{H_*} \left(\frac{T_{*}}{100 \text{ GeV}}\right)\left(\frac{g_*}{100} \right)^{\frac{1}{6}} \text{ Hz},
\end{equation}
which is associated with the following energy density
\begin{equation}
     h^2\Omega_{\text{col}} (f) = 1.67 \times 10^{-5} \left( \frac{H_*}{\beta}\right)^{2} \frac{0.11 v_w^3}{0.42+v_w^2} \left(\frac{\kappa \alpha}{1+\alpha}\right)^2 \left(\frac{100}{g_*}\right)^{\frac{1}{3}} \frac{3.8 \left(f/f_{\text{col}}\right)^{2.8}}{1+2.8 \left(f/f_{\text{col}}\right)^{3.8}},
\end{equation}
where \(g_*\) is the number of relativistic degrees of freedom at temperature \(T_*\), \(\kappa\) is the efficiency factor given by
\begin{equation}
    \kappa = \frac{\alpha_{\infty}}{\alpha} \left (\frac{\alpha_{\infty}}{0.73+0.083 \sqrt{\alpha_{\infty}}+\alpha_{\infty}} \right),
\end{equation}
and the velocity of the bubble wall reads
\begin{equation}\label{velocity}
    v_w = \frac{1/\sqrt{3}+ \sqrt{\alpha^2+2 \alpha/3}}{1+\alpha}.
\end{equation}
According to the approach above \cite{Espinosa:2010hh, Bodeker:2009qy}, the energy, which is transferred into the fluid, saturates at 
\begin{equation}
    \alpha_{\infty} = 0.49 \times 10^{-3} \left(\frac{\upsilon_*}{T_*} \right)^2
\end{equation}
and the bubble wall velocity (\ref{velocity}) offers only a lower bound on the actual velocity of the bubble wall \cite{Huber:2008hg}. Next, the contribution of the sound waves to the spectrum of gravitational waves, after the phase transition has completed, is given by \cite{Hindmarsh:2013xza, Hindmarsh:2015qta}
\begin{equation}
   h^2 \Omega_{\text{sw}} (f) = 2.65 \times 10^{-6} \left(\frac{H_{*}}{\beta} \right) \left( \frac{\kappa \alpha}{1+\alpha}\right)^2 \left(\frac{100}{g_{*}}\right)^{\frac{1}{3}} v_w \left(\frac{f}{f_{\text{sw}}}\right)^3 \left(\frac{7}{4+3 \left(f/f_{\text{sw}}\right)^2}\right)^{\frac{7}{2}},
\end{equation}
where the frequency at the peak of this contribution is computed as
\begin{equation}
    f_{\text{sw}} = 1.9 \times 10^{-5} \frac{\beta}{H} \frac{1}{v_w} \left(\frac{T_{*}}{100 \text{ GeV}}\right) \left(\frac{g_{*}}{100}\right)^{\frac{1}{6}} \text{ Hz}.
\end{equation}
Finally, the MHD turbulence contributes to the gravitational wave background with the following energy density \cite{Binetruy:2012ze}, 
\begin{equation}
    h^2 \Omega_{\text{turb}}  (f) = 3.35 \times 10^{-4} \left(\frac{H_{*}}{\beta} \right) \left( \frac{\epsilon \kappa \alpha}{1+\alpha}\right)^{\frac{3}{2}} \left(\frac{100}{g_{*}} \right)^{\frac{1}{3}} v_w \frac{\left( f/f_{\text{turb}}\right)^3\left(1+f/f_{\text{turb}}\right)^{-\frac{11}{3}}}{1+8 \pi f a_{0}/\left(a_{*} H_{*}\right)},
\end{equation}
where \(\epsilon \simeq 0.05\) and the corresponding peak frequency reads
\begin{equation}
    f_{\text{turb}} = 2.7 \times 10^{-5} \left(\frac{\beta}{H_{*}}\right) \frac{1}{v_w} \left(\frac{T_{*}}{100 \text{ GeV}}\right) \left(\frac{g_{*}}{100} \right)^{\frac{1}{6}} \text{Hz}.
\end{equation}
In general, one observes that the magnitude of the gravitational wave signals increases with the strength of the phase transition \(\left(\upsilon_c/T_c\right)\) \cite{Beniwal:2017eik}. This implies that the majority of the parameter space of the real singlet extension, which satisfies the condition for a successful EWBG, is also associated with an observable gravitational wave signal. As a result, if the real singlet extension describes a successful EWBG, gravitational waves are generated due to the electroweak phase transition that can be detected by the future gravitational wave interferometers, such as LISA \cite{Bartolo:2016ami}.

\end{document}